\pdfoutput=1
\documentclass[12pt,english,american,preprint]{article}
\usepackage[latin9]{inputenc}
\usepackage{color}
\usepackage{array}
\usepackage{float}
\usepackage{multirow}
\usepackage{amsmath}
\usepackage{amssymb}
\usepackage{graphicx}
\usepackage{esint}

\makeatletter

\providecommand{\tabularnewline}{\\}

\usepackage{graphics}\usepackage{epsfig}\usepackage{amsfonts}
\renewcommand{\vec}[1]{{\bf #1}}
\newcommand{\eqb}{\begin{equation}}
\newcommand{\eqe}{\end{equation}}
\newcommand{\dmb}{\begin{displaymath}}
\newcommand{\dme}{\end{displaymath}}

\newcommand{\eab}{\begin{eqnarray}}
\newcommand{\eae}{\end{eqnarray}}
\newcommand{\ra}{\rangle}
\newcommand{\la}{\langle}
\newcommand{\e}{\mbox{e}}
\newcommand{\be}{\begin{equation}}
\newcommand{\ee}{\end{equation}}

\usepackage{slashbox} 

\makeatother

\usepackage{babel}
\begin{document}
\begin{titlepage} 

\vspace{0.6cm}

\begin{center}
{\Large {One-loop photon-photon scattering in a thermal, deconfining
SU(2) Yang-Mills plasma} \vspace{1.5cm}
}\\
{\Large{} }{\large {Niko Krasowski and Ralf Hofmann$^*$} }
\par\end{center}

\vspace{0.5cm}

\begin{center}
Institut f\"ur Theoretische Physik\\
Universit\"at Heidelberg\\
Philosophenweg 16\\
69120 Heidelberg, Germany
\end{center}

\vspace{0.5cm}
 
\begin{abstract}
For a deconfining thermal SU(2) Yang-Mills plasma we discuss
the role of (anti)calorons in introducing non-thermal behavior effectively 
described in terms of Planck's quantum of action $\hbar$. 
This non-thermality cancels exactly between the ground-state
estimate and its free quasiparticle excitations. Kinematic constraints in 4-vertex 
scattering and the counting of radial loop variables versus the number of independent constraints on them 
are re-visited. Next, we consider thermal 2$\rightarrow$
2 one-loop scattering of the modes remaining massless upon the
(anti)caloron induced adjoint Higgs mechanism (thermal ground state
after spatial coarse graining). Starting
with stringent analytical arguments, we are able to exclude 
the contribution to photon-photon scattering from diagrams containing at least 
one three-vertex and, in a next step, a vast majority of all 
possible configurations involving two four-vertices. By numerical analysis we 
show that the remaining contribution of the overall S channel is severely suppressed compared to that of the T and U channels, meaning 
that the creation of a pair of massive vector modes by a pair of photons and 
vice versa practically does not occur in the Yang-Mills plasma. For the T and U channels the domain of loop integration represents  
less than $10^{-7}$ times the volume of the 
unconstrained integration region. The thus introduced photon-photon correlation should affect  
the Cosmic Microwave Background's polarisation 
at low redshift. 
An adaption of the here-developed methods to the analysis of 
irreducible bubble diagrams could prove the conjecture of 
hep-th/0609033 on the termination of the loop expansion of thermodynamical quantities 
at a finite irreducible order.      
\end{abstract}$\mbox{}$\vspace{1.0cm}\\ 

$^*$email: r.hofmann@thphys.uni-heidelberg.de
\end{titlepage}

\section{Introduction}

Effective, indeterministic behavior, inherent to scattering amplitudes in quantum
Yang-Mills theory, appears to be associated with the presence of \textsl{classical}
field configurations of finite action and nontrivial topology in 4D
Euclidean spacetime \cite{BPST1975,Jackiw1976,'tHooft1976U,'tHooft1976}. 
For SU(2) Yang-Mills thermodynamics a concrete argument in favor of 
this idea was put forward recently \cite{KavianiHofmann2012,Hofmann2012}.
Namely, in its deconfining phase an adjoint Higgs mechanism, caused
by calorons and anicalorons of topological charge modulus unity, which, upon spatial
coarse graining \cite{HerbstHofmann2004}, effectively materialize in terms of an inert scalar field $\phi$, 
splits the Yang-Mills mass spectrum
into two degenerate  modes propagating on their quasiparticle mass-shell (unitary
gauge) and a massless mode (photon) whose off-shellness is constrained
by $|\phi|$ (Coulomb gauge) \cite{Hofmann2005}. Moreover, (anti)caloron
mediated 4-vertices (and 3-vertices) are pointlike in the effective theory because
four-momentum transfer through them is restricted by
$|\phi|$ in all Mandelstam variables. The present paper intents to investigates the consequences
of these constraints for photon-photon scattering in the effective theory for the deconfining phase. 
However, before performing the according analysis, we would like to elaborate on the
argument for and the implied consequences of (anti)caloron mediation of scattering in effective
Yang-Mills vertices. To do so, we work in units where $c$, the speed of light in
vacuum, is set to unity but Boltzmann's constant $k_{B}$ and Planck's
quantum of action $\hbar$ are dimensionful.

Due to Feynman \cite{Feynman1953,FeynmanHibbs1965} and Schwinger
\cite{Migdal2006} the fundamental Yang-Mills partition
function $Z$ is representable in fundamental variables as follows
\begin{align}
Z\equiv\mbox{tr }e^{-\frac{\beta}{k_{B}}\mathcal{H}}=\sum_{n}\langle\left.n\right|\mbox{tr }e^{-\frac{\beta}{k_{B}}\mathcal{H}}\left|n\right.\rangle=
\int_{A_{\mu}\left(0,\mathbf{x}\right)=A_{\mu}\left(\beta,\mathbf{x}\right)}\mathcal{D}A\, \e^{-\frac{1}{k_{B}}\int_{0}^{\beta}d\tau^\prime\, 
d^{3}x\,\mathcal{L}_{E}\left[A_{\mu}\right]}\,.\label{partFunc}
\end{align}
where $\beta\equiv T^{-1}$, $T$ denotes temperature, $\mathcal{H}$ represents the Hamiltonian of quantum Yang-Mills theory, 
the functional integration is over gauge-inequivalent, periodic field configurations, and 
$\mathcal{L}_{E}$ refers to the classical Euclidean action density. In evaluating partition function (\ref{partFunc}), 
a simplification occurs if the quantum states $|n\rangle$ are taken
to be (gauge-inequivalent) energy eigenstates of ${\cal H}$ with eigenvalues $E_{m}$. Formally, the partition function then reads
\begin{align}
Z=\sum_{m}M(m)e^{-\frac{\beta E_m}{k_B}}\,.\label{partfuncteigH}
\end{align}
Here $M(m)$ denotes the degeneracy of $E_{m}$. Resorting to a represention
of the Hilbert space in field variables, ${\cal H}$ is a gauge-invariant
functional of gauge field $A_{\mu}$ and its conjugate momentum
$\Pi_{\mu}$ whose quantization conditions (equal-time commutators
of $A_{\mu}$ and $\Pi_{\mu}$) involve $\hbar$. In a Euclidean spacetime,
the operator $\e^{-\frac{\beta}{k_{B}}{\cal H}}$ formally evolves
a state $|n\ra$ in $\tau^{\prime}$ -- a variable of dimension inverse
temperature $T^{-1}$ -- from $\tau^{\prime}=0$ to $\tau^{\prime}=\beta\equiv 1/T$
if $\tau^{\prime}$ and time $\tau$ are related as 
\eqb 
\label{tautauprime} 
\tau^\prime=\frac{k_B}{\hbar}\tau\ \ \ \ \mbox{or}
\ \ \ \ \tau=\frac{\hbar}{k_B}\tau^\prime\,. 
\eqe 
Now, \textsl{neither the macroscopic concept time ($\tau$)
nor the macroscopic concept temperature ($\tau^{\prime}$) ought to depend
on the values of $\hbar$ and $k_{B}$}. As a consequence, Eq.\,(\ref{tautauprime})
implies that in the formal limit $\hbar\to 0$ also $k_{B}\to 0$ such
that 
\eqb
\label{hbarproptokB} 
\frac{\hbar}{k_B}=\mbox{const}\,.
\eqe 
Quantity $\la n|\e^{-\frac{\beta}{k_{B}}{\cal H}}|n\ra$ thus represents
the formal persistence amplitude for state $|n\ra$ under (Euclidean)
time evolution from $\tau=0$ to $\tau=\frac{\hbar}{k_{B}T}$. 

If the world were purely thermal and Euclidean (no reference
to a Wick rotated version of temporal evolution acting on a quantum state in
Minkowski spacetime) on subatomic length scales then the concept of action would be inappropriate, and 
energy per temperature in units of $k_{B}$ would occur naturally in 
exponential weights representing the Yang-Mills spectrum. Conversely, if the subatomic world were purely Minkowskian
and non-thermal then one would consider a summation over field `trajectories' whose action in units of $\hbar$ weights their contributions to 
the temporal evolution of a given quantum state. Interestingly, thermal Quantum Yang-Mills theory emerges as an interplay of these two 
concepts. For example, the notion of the thermal ground state invokes the fact 
that the former point of view is related to the latter by representing effective variables nonlocally in terms of fundamental 
ones: The periodicity of these fundamental (classical, Euclidean) configurations (calorons and anticalorons) is a consequence of 
identification (\ref{tautauprime}) which rests on temporal evolution of quantum states. Collectively, however, 
nontrivial, temporal periodicity collapses to a mere choice of gauge on the effective-theory level \cite{Hofmann2005}. 
Alternatively, the occurrence of effective quantum corrections, expanded 
in powers of $\hbar$, is demanded by isolated action of (anti)calorons. In this context, hinging on adiabatic slowness 
to render Euclidean signature irrelevant, the 
physics of (anti)caloron constituents (monopole-antimonopole pair) can be interpreted classically \cite{Diakonov2004,Ludescheretal2010}.  

For the deconfining phase of SU(2) Yang-Mills thermodynamics we have
in \cite{KavianiHofmann2012,Hofmann2012} identified the Euclidean
action of relevant calorons or anticalorons \cite{HS1977,Nahm1983,VanBaal1998,LeeLu1998}
with $\hbar$. The according chain of arguments relies on a Minkowskian
\textsl{interpretation} of free but effectively massive (adjoint Higgs
mechanism) one-loop fluctuations about the ground-state estimate.
Here the contribution of quasiparticles to the pressure, expressed
by a Matsubara sum in Euclidean spacetime, is cast into a continuous
integration by employing Cauchy's integral theorem. Because of Minkowskian
on-shellness (inertness of the adjoint scalar field $\phi$ \cite{Hofmann2011}) only the Bose
weighted part in the propagator contributes to the loop integration. Thermodynamic consistency
(linking of thermodynamical quantities by Legendre transformations
motivated by the left part of Eq.\,(\ref{partFunc})) and a dimensional
counting in effective interaction monomials with appropriately normalized
field variables \cite{Brodsky2010} determines the effective coupling
$e$ in units of $\hbar^{-1/2}$ \cite{KavianiHofmann2012,Hofmann2012}.
Finally, one appeals to the observation that (anti)caloron radii $\rho\sim|\phi|^{-1}$
dominate the spatial coarse-graining -- generating the field $\phi$ \cite{HerbstHofmann2004,Hofmann2011} -- to justify
the use of $e$ in the formula $S=\frac{8\pi^{2}}{e^{2}}$ for the
Euclidean action of the fundamental field configurations
caloron or anticaloron with topological charge modulus unity\footnote{It was shown in \cite{HerbstHofmann2004} that 
(anti)calorons of higher charge modulus do not contribute to $\phi$.}. 
As an immediate consequence, pointlike effective vertices, 
introducing perturbations into thermally weighted Minkowskian on-shell propagation in terms of free plane waves, 
occur because these vertices effectively interpolate
the incoming and the scattered wave through spatially unresolved (anti)calorons
subject to a \textsl{Euclidean} time dependence. Here, the irreconcilability
of Euclidean and Minkowskian time evolution \textsl{within} a vertex, represents 
the cause of effective indeterminism in the outcome of the scattering
event: Wick rotation of a caloron's time dependence
generates a field configuration which is far from the nearest classical
solution to the Minkowskian Yang-Mills equations. Effectively parametrized
by $\hbar$, the action of isolated (anti)calorons thus introduces
disorder to equilibrated Minkowskian plane-wave
propagation\footnote{This equilibration is by interaction with the 
ground-state estimate which, in turn, invokes a sum over an infinite series 
of effective forward-scatterings induced by (anti)calorons: the emergence of 
quasiparticle mass by the adjoint Higgs mechanism. Notice that 
the Bose-Einstein function does not depend on $\hbar$ separately but on
the ratio $\frac{\hbar}{k_{B}}$, see discussion of Eq.\,(\ref{hbarproptokB})
above.}. In this sense (anti)calorons behave non-thermally, resolving the
following paradox associated with the ground-state estimate: While the latter, describing 
the interacting (anti)calorons in terms
of the field $\phi$ and an effective, pure-gauge configuration $a_{\mu}^{\tiny\mbox{gs}}$, 
obeys an equation of state $P^{\tiny\mbox{gs}}=-\rho^{\tiny\mbox{gs}}$
it does exhibit a finite heat capacity. In other words, the ground state's
entropy density $s^{\tiny\mbox{gs}}$, given as 
\eqb 
\label{entropydens} s{\tiny \mbox{gs}}\equiv\frac{1}{T}(P{\tiny \mbox{gs}}+\rho{\tiny \mbox{gs}})\,,
\eqe 
vanishes although $\rho^{\tiny\mbox{gs}}=4\pi T\Lambda^{3}$ is temperature
dependent, $\Lambda$ denoting the Yang-Mills scale \cite{Hofmann2005}. At finite temperature this violates  
the third law of thermodynamics. Therefore, a thermodynamical interpretation of the ground-state
physics alone is not admissible. That this estimate does not behave 
thermodynamically by itself is, however, not a surprise because of the 
de-thermalizing effect of (anti)calorons constituting the thermal ground state. Modulo small radiative corrections, see below, the overall Yang-Mills
system is thermodynamical, however: between the ground
state and its quasiparticle excitations the $T$ dependence of the effective coupling
$e$ cancels non-thermality \cite{Hofmann2007}. When, at the critical temperature $T_{c^{\prime}}$ 
(lowest attainable temperature in the preconfining phase \cite{Hofmann2005}), quasiparticles disappear entirely 
the thermal ground state would represent the entire thermal partition function which, by the above entropy argument, does not make sense. 
Indeed, the onset of the Hagedorn transition at $T_{c^{\prime}}$ explicitly violates a thermodynamical description 
of the Yang-Mills system \cite{Hofmann2007}. Note that except for the limit $T\searrow T_{c}$ \cite{Hofmann2005}, where
radiative corrections vanish \cite{Hofmann2005}, non-thermal effects
are ubiquitous also in the deconfining and preconfining phases: there are small
radiative corrections in the former phase, associated with fixed or resummed loop orders \cite{SchwarzHofmannGiacosa2007,LudescherHofmann2009,FalquezHofmann2010} 
which are triggered by isolated (anti)calorons, while supercooling and a related ground-state tunneling take place in the 
latter phase \cite{Hofmann2007}.

To conclude this discussion, we may simply state that the (deeply physical) demand for periodicity of {\sl any} field 
configuration contributing to the reformulation of the partition function in terms of the 
functional integral in Eq.\,(\ref{partFunc}) is an extrapolation of the 
free-field case which introduces in principle non-thermal behavior: While the Minkowskian on-shell propagation of 
free plane waves becomes Bose-Einstein weighted as a consequence of this prescription it also allows 
for the contribution of topologically charged field configurations (calorons and anticalorons) 
which ultimately (low-temperature situation) 
destroy spatial translation invariance and which introduce an over-exponentially rising density of 
states in the confining phase. As long as the collective effect of (anti)calorons gives 
rise to spatially homogeneous effects only (e.g., energy density of the ground state or 
quasiparticle masses) thermodynamics is kept intact by cancellations of non-thermal 
effects between ground state and massive excitations, enabled by a proper 
tuning of the coupling (deconfining phases). Radiative 
corrections, however, {\sl are} mediated by {\sl isolated} 
(anti)caloron action, altering the propagation behavior 
of every single plane wave in a non-thermal way. We believe that this is the 
cause for the large-angle anomalies in the temperature fluctuations of the Cosmic 
Microwave Background presently discussed.             

The present paper is organized as follows. In Sec.\,\ref{ReT} we review
the effective theory together with its constraints in unitary-Coulomb
gauge, and we re-revisit the counting of constraints in dependence on loop order in irreducible 
diagrams subject to 4-vertices only. Sec.\,\ref{Pps} discusses how these constraints apply 
to exclude certain one-loop diagrams potentially contributing to the photon-photon scattering amplitude. Moreover, 
a technology is developed that allows to systematically discriminate a number of 
channel combinations for the two 4-vertices in the remaining diagrams in this amplitude. 
For the remaining possibilities, where no further analytical assessment
can be made, a Monte-Carlo simulation of the domain of integration
is performed, and a selected hit is investigated further in view of its 
embedding into the extremely fillamented algebraic variety 
comprising a part of the domain of integration. The last section summarizes our results 
and gives conclusions.

\section{Effective theory, notational conventions, and constraints\label{ReT}}

For the benefit of the reader we re-visit briefly the
effective theory for deconfining SU(2) Yang-Mills thermodynamics and
discuss its constraints on 4-vertices, see also \cite{Hofmann2012,Hofmann2011}. We also discuss an 
(inessential) correction to the counting of constraints for irreducible diagrams that arise 
solely from 4-vertices. From now on we work in supernatural units $\hbar=k_{B}=c=1$.

\subsection{Effective action}

The effective action for the deconfining phase emerges upon a spatial
coarse-graining over free, trivial-holonomy (anti)calorons of charge modulus unity and plane-wave
fluctuations \cite{HerbstHofmann2004}. While the topologically nontrivial
sector of gauge-field configurations yields an inert, adjoint scalar field
$\phi$ coarse-graining over interactions within the topologically
trivial sector is determined by perturbative renormalizability\footnote{Because, a priori these interactions between fundamental modes are
characterized by momentum transfers $>|\phi|^{2}$ loop momenta in these fundamental fluctuations are
far off their mass shell. Thus no thermal treatment is required in
the formal coarse-graining over the topologically trivial sector.
Moreover, the subtraction of ultraviolet divergences in perturbation
theory is consistent with the fact that calorons of radius $\rho\sim|\phi|^{-1}$
sharply dominate the coarse graining. Thus only a thin shell in momentum transfer can be mediated by them 
\cite{KavianiHofmann2012,Hofmann2012,HerbstHofmann2004}.} 
to contribute to the effective action for the coarse-grained,
topologically trivial gauge field $a_{\mu}$ in the same
form as the fundamental Yang-Mills action \cite{'tHooftVeltman,ZinnJustin}.
Moreover, the appearance of mixed operators of mass dimension higher
than four, involving both fields $a_{\mu}$ and $\phi$, is excluded
because momentum transfer to the field $\phi$ is impossible \cite{Hofmann2011}. 
Thus the unique (Euclidean) density of effective action ${\cal L}_{\mbox{\tiny eff}}[a_{\mu}]$ is given as
\begin{equation}
{\cal L}_{\mbox{\tiny eff}}[a_{\mu}]=\mbox{tr}\,\left(\frac{1}{2}\, 
G_{\mu\nu}G_{\mu\nu}+(D_{\mu}\phi)^{2}+\frac{\Lambda^{6}}{\phi^{2}}\right)\,,\label{fullactden}
\end{equation}
where $G_{\mu\nu}=\partial_{\mu}a_{\nu}-\partial_{\nu}a_{\mu}-ie[a_{\mu},a_{\nu}]\equiv G_{\mu\nu}^{a}\, t_{a}$
denotes the field strength, $D_{\mu}\phi=\partial_{\mu}\phi-ie[a_{\mu},\phi]$,
and $e$ is the effective gauge coupling yet to be determined. ${\cal L}_{\mbox{\tiny eff}}$
in Eq.\,(\ref{fullactden}) yields a highly accurate tree-level ground-state
estimate and, as easily deduced in unitary gauge $\phi=2|\phi|\, t_{3}$,
a tree-level mass $m=2\, e|\phi|=2\, e\sqrt{\frac{\Lambda^{3}}{2\pi T}}$
for propagating gauge modes $a_{\mu}^{1,2}$. Because the mass of
these modes is generated by a summation of a Dyson series of up to
infinitely many local interactions with the field $\phi^{2}$ the off-shellness
inducing absorption or emission of effective, propagating modes is forbidden: these processes 
would introduce momentum transfer to the inert field
$\phi$. As a consequence, massive modes propagate thermally on their
mass shell \cite{Hofmann2011}. Massless modes (photons), on the other
hand, may be off their mass shell at most by $|\phi|^{2}$ \cite{Hofmann2005}.
Finally, thermodynamical consistency of the system's ground-state estimate and its free 
quasiparticles yields the value $e=\sqrt{8}\pi$ almost everywhere. The exception is a narrow, logarithmic 
pole at $T_{c}=13.87\,\frac{\Lambda}{2\pi}$.

\begin{figure}[h]
\centering{}\includegraphics[width=1\textwidth]{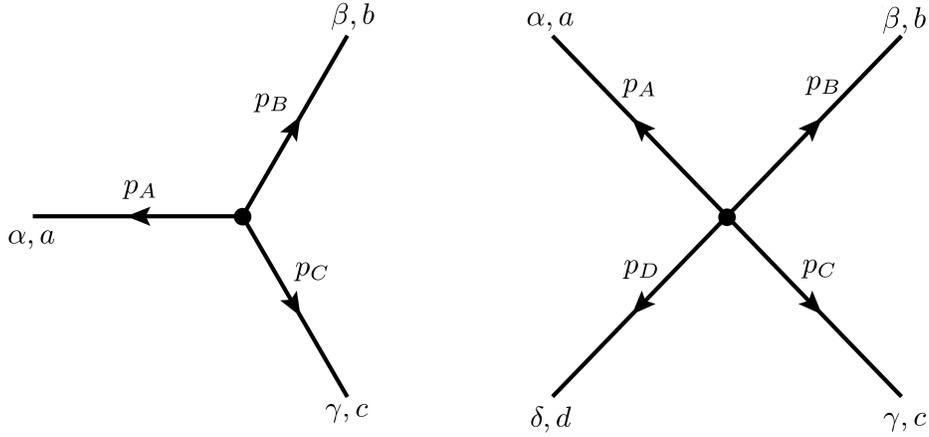}
\caption{Yang-Mills vertices, $p_{X}$ being a four-momentum, greek letters
are Lorentz indices, lower case latin letters denote Lie-Algebra indices.\label{gen_vert}}
\end{figure}

Considering radiative corrections, the effective 3- and 4-vertices
are given as 
\begin{align}
\Gamma_{[3]\, abc}^{\alpha\beta\gamma}= & \, e(2\pi)^{4}\delta(p_{A}+p_{B}+p_{C})\epsilon_{abc}\times\left[g^{\alpha\beta}\left(p_{B}-p_{A}\right)^{\gamma}\right.\nonumber \\
 & \left.+g^{\beta\gamma}\left(p_{C}-p_{B}\right)^{\alpha}+g^{\gamma\alpha}\left(p_{A}-p_{C}\right)^{\beta}\right]\,,\label{eq:AllgemeineFormel3Vertex}\\
\nonumber \\
\Gamma_{[4]\, abcd}^{\alpha\beta\gamma\delta}= & -\mathrm{i}e^{2}(2\pi)^{4}\delta(p_{A}+p_{B}+p_{C}+p_{D})\times\left[\epsilon_{abe}\epsilon_{cde}(g^{\alpha\gamma}g^{\beta\delta}-g^{\alpha\delta}g^{\beta\gamma})\right.\nonumber \\
 & \left.+\epsilon_{ace}\epsilon_{bde}(g^{\alpha\beta}g^{\gamma\delta}-g^{\alpha\delta}g^{\beta\gamma})+\epsilon_{ade}\epsilon_{bce}(g^{\alpha\beta}g^{\gamma\delta}-g^{\alpha\gamma}g^{\beta\delta})\right]\,,\label{eq:AllgemeineFormel4Vertex}
\end{align}
compare with Fig.\,\ref{gen_vert}. 
In the present paper, no explicit reference to propagators is made
except for the above mentioned fact that massive modes always are on-shell. 
Let us now set up our notational conventions.\vspace{0.2cm}
\\
{\sl Vertices.} Vertices are discussed on different
levels. Some arguments require only formulas that are valid independently
of the sort of attached gauge modes. In such cases we just use the
symbols $p_{1}$, $p_{2}$ etc. for four-momenta. On a level, where
we would like to distinguish between massive and massless modes, we
use $R$, $S$, $P$ for the four-momenta of the massive modes
and $p$, $q$ for the four-momenta of photons. In applications
to the actual scattering process we employ $a$ and $b$ to indicate
the four-momenta of incoming photons. Outgoing photons are labeled
by $c$ and $d$. Four-momenta of internal, massive (loop) modes are
denoted by $u$ and $v$.\vspace{0.2cm}
\\
{\sl Feynman diagrams.} In general, the propagation of massive particles
is indicated by double lines in Feynman diagrams while massless modes
are represented by single lines (except for Fig.\,2 where a wavy line is used). \vspace{0.2cm}
\\
{\sl Scattering channels.} An overall channel in photon-photon scattering
is labeled by the Mandelstam variable S,T or U (captial letters). On the
other hand, for scattering channels associated with a given 4-vertex
Mandelstam variables are in lower case letters. In this way, one specific
configuration can be written in a short-hand notation. For example,
``Stu'' describes overall S-channel scattering with the t-channel
realized at the first 4-vertex and the u-channel at the second 4-vertex.

\subsection{3-vertex}

\begin{figure}[h]
\begin{centering}
\includegraphics[width=0.5\textwidth]{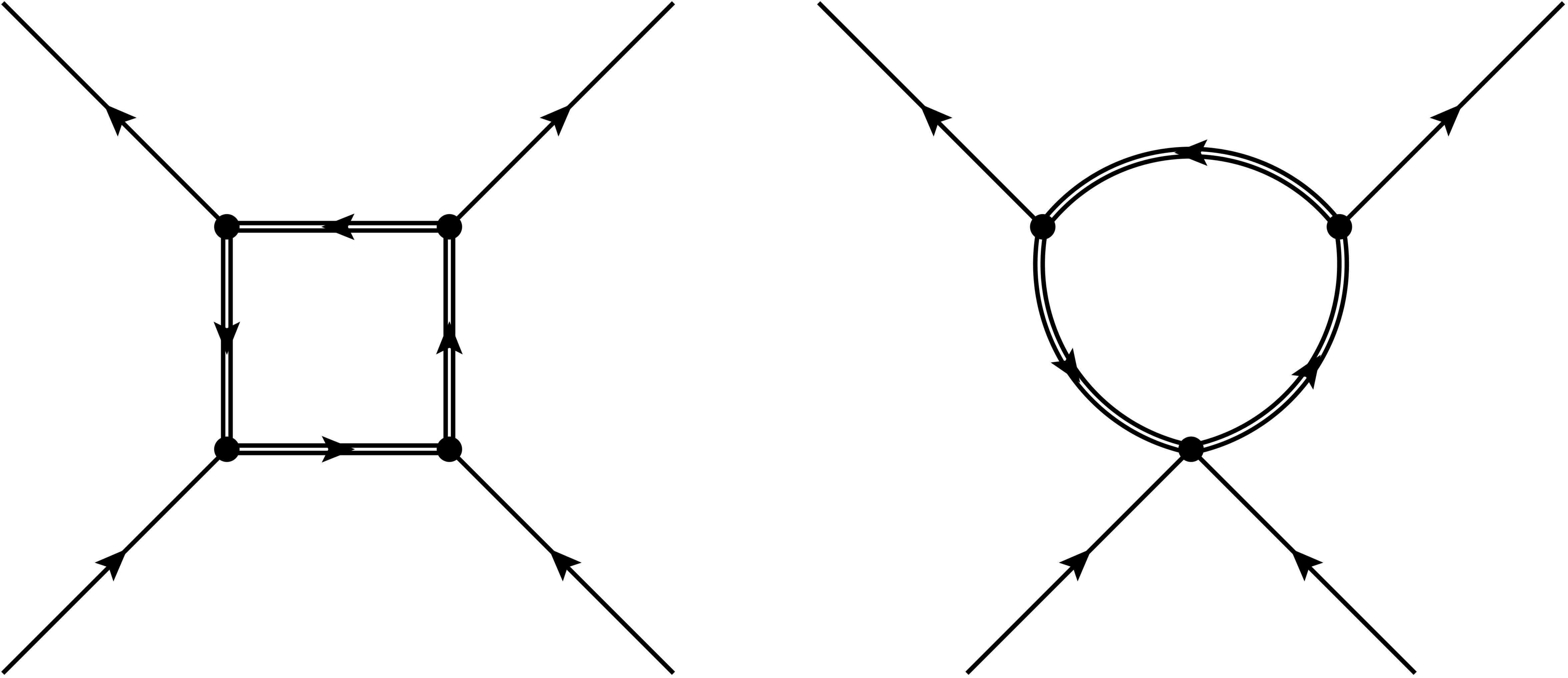} 
\par\end{centering}

\caption{Box diagram (left) and penguin diagram (right). Both possibilities
are excluded to contribute to photon-photon scattering. \label{boxandping}}
\end{figure}

\noindent At the 3-vertex, given as in Eq.\,(\ref{eq:AllgemeineFormel3Vertex}), a photon always connects to two massive modes 
\footnote{This is due to the Levi-Civita symbol in Eq.\,(\ref{eq:AllgemeineFormel3Vertex}).}. 
\begin{figure}[h]
\begin{centering}
\includegraphics[width=0.5\textwidth]{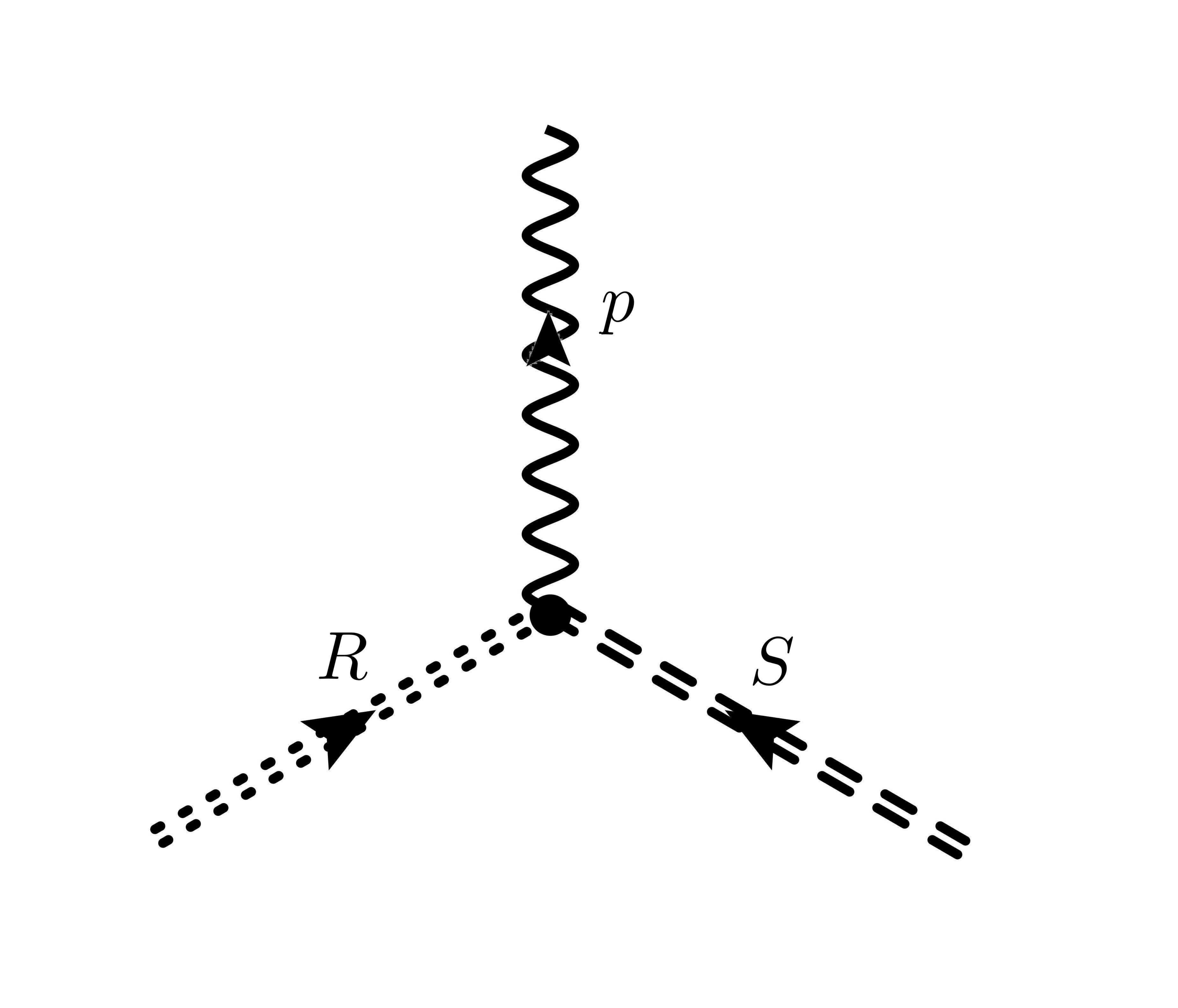} 
\par\end{centering}
\caption{Two different massive modes (dashed and dotted) and a massless one
(wavy) at a 3-vertex.}
\end{figure}
For later use let us now check whether the photon can be an external, on-shell particle of positive energy 
\footnote{\noindent In the present paper we neglect modifications of the free
dispersion law $p_{0}=|\vec{p}|$ due to a resummation of insertions
of the one-loop polarization tensor, see \cite{SchwarzHofmannGiacosa2007,LudescherHofmann2009}.
These effects are sizable at low energy and temperatures, and they
die off in an exponential way in the former and a power-like way in
the latter variable.}. Because of four-momentum conservation and the on-shellness of the
massive modes the following conditions apply: 
\begin{align}
R^{2}=S^{2} & =m^{2}\,,\nonumber \\
p^{2} & =0\,,\nonumber \\ 
(R+S)^{2} & =p^{2}=0\,.
\end{align}
The energy of the massive particles can be positive or negative, depending
on the direction of energy flow in the loop of the overal scattering
diagram, see Fig.\,\ref{boxandping}. 
Therefore, one has 
\begin{align}
2m^{2}\pm2\sqrt{\left|\mathbf{R}\right|+m^{2}}\sqrt{\left|\mathbf{S}\right|^{2}+m^{2}}-2\mathbf{R}\mathbf{S} & =0\nonumber \\
{\mbox{upon squaring}}\,\Rightarrow\;\;\left(\left|\mathbf{R}\right|^{2}+m^{2}\right)\left(\left|\mathbf{S}\right|^{2}+m^{2}\right)-m^{4}-
\left(\mathbf{R}\mathbf{S}\right)^{2}+2m^{2}\mathbf{R}\mathbf{S} & =0\nonumber \\
 \Rightarrow\;\;m^{2}\left(\left|\mathbf{R}\right|^{2}+\left|\mathbf{S}\right|^{2}+2\left|\mathbf{R}\right|\left|\mathbf{S}\right|
\cos\measuredangle(\mathbf{R},\mathbf{S})\right)\nonumber \\ 
+|\mathbf{R}|^{^{2}}\left|\mathbf{S}\right|^{2}
\left(1-\cos^{2}\measuredangle(\mathbf{R},\mathbf{S})\right) & =0\,.\label{3vertmomcon}
\end{align}
\\
The minimal value of each summand in the last equation of \eqref{3vertmomcon}
is zero: The first summand vanishes for $\cos\measuredangle(\mathbf{R},\mathbf{S})=-1$
and $\left|\mathbf{R}\right|=\left|\mathbf{S}\right|$, the second
summand is zero for $\cos\measuredangle(\mathbf{R},\mathbf{S})=\pm1$.
So the only configuration satisfying Eq.\,(\ref{3vertmomcon})
is $\measuredangle(\mathbf{R},\mathbf{S})=\pi$ and equal energy of
the photons. Since in a loop integral the thus allowed integration
over angular variables is over a hypersurface of measure zero we conclude
that 3-vertices do not contribute to the overall one-loop scattering
of photons.

\subsection{Constraints on 4-vertex\label{4-vertex}}

\begin{figure}[H]
\begin{centering}
\includegraphics[width=0.3\textwidth]{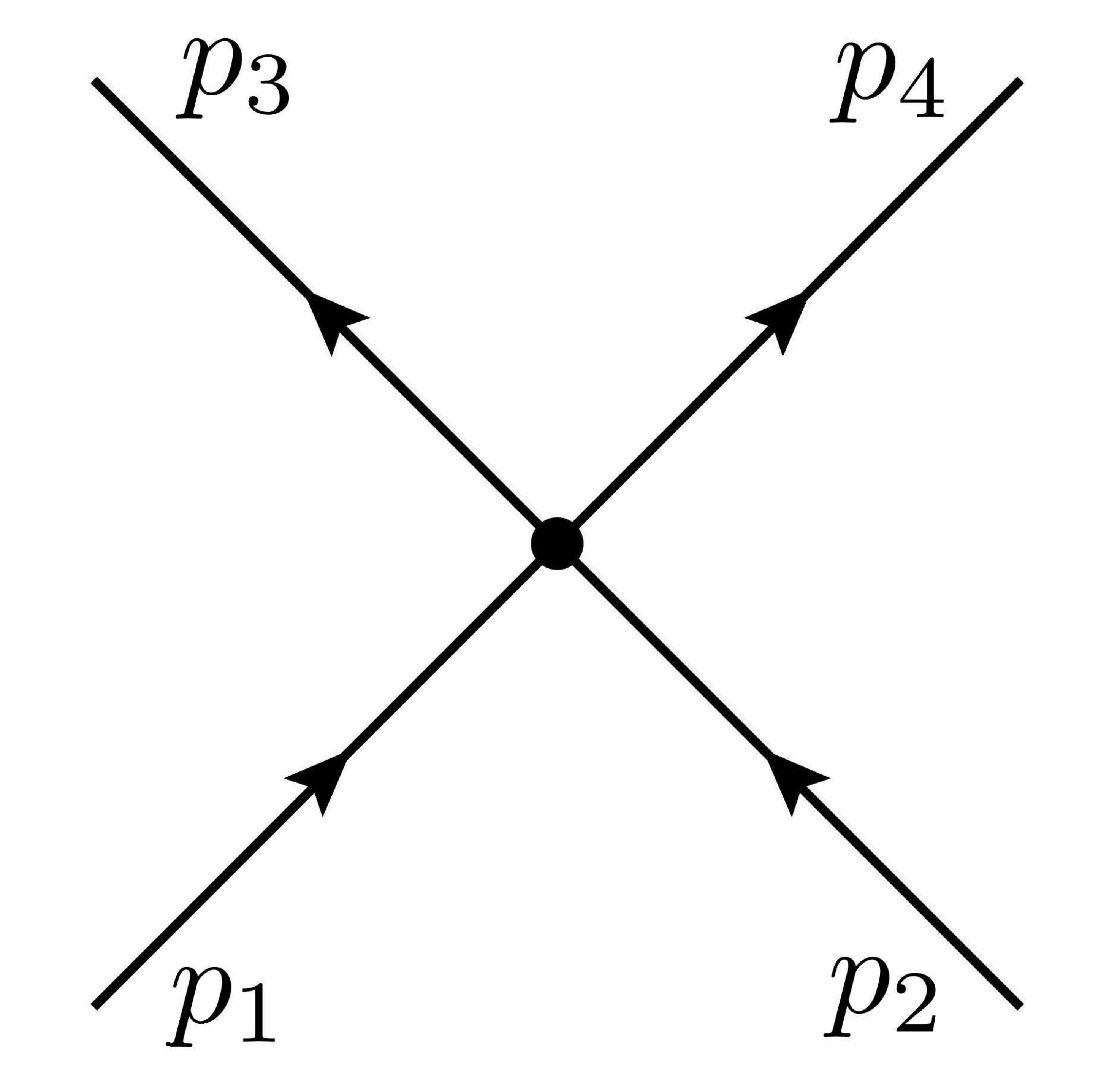} 
\par\end{centering}
\caption{Scattering due to a 4-vertex.\label{Gen4scat}}
\end{figure}

It is easy to show that the 4-vertex is invariant under a permutation
of the legs attached to it \cite{Krasowski2012}. This implies
that the 4-vertex does not distinguish between Mandelstam variables
$s,t,$ and $u$ in mediating 2$\to$2 scattering. The according constraints,
however, do. It is also straightforward to demonstrate that, modulo leg permutation, only the 
two diagrams depicted in Fig.\,\ref{fig:Massless,-massless-to hh u hh to hh}
take place. 
\begin{figure}[H]
\begin{centering}
\includegraphics[width=0.8\textwidth]{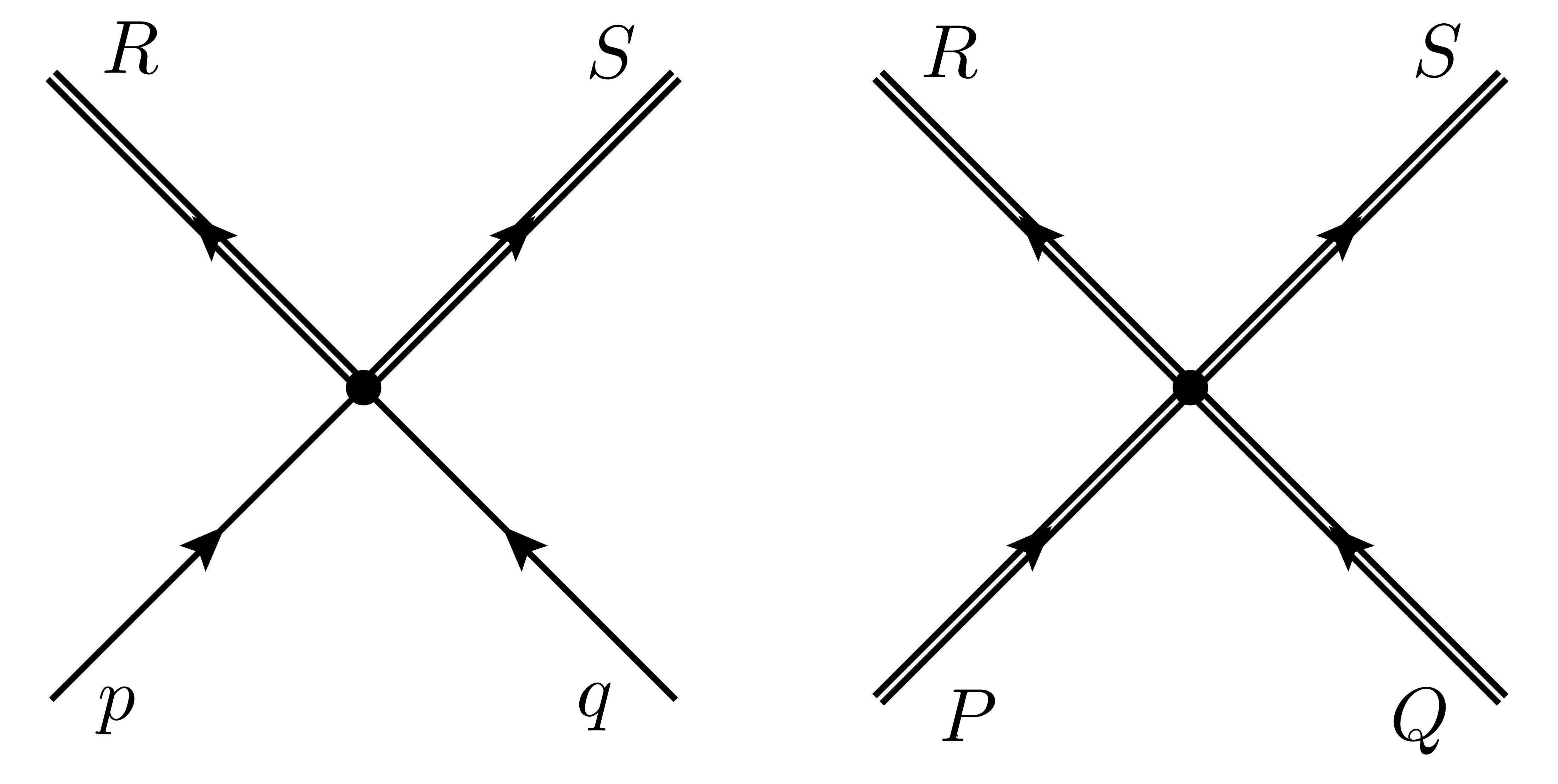} 
\par\end{centering}
\caption{\label{fig:Massless,-massless-to hh u hh to hh}Photon-photon to massive-massive
and massive-massive to massive-massive scattering. These are the cases
that need to be distinguished in all scattering channels.}
\end{figure}
\noindent In the effective theory one has \cite{Hofmann2005}
\begin{align}
s=\left|\left(p_{1}+p_{2}\right)^{2}\right|=\left|\left(p_{3}+p_{4}\right)^{2}\right|\leq\left|\phi\right|^{2}\,,\label{eq:VertexChConstraints}\\
t=\left|\left(p_{1}-p_{3}\right)^{2}\right|=\left|\left(p_{2}-p_{4}\right)^{2}\right|\leq\left|\phi\right|^{2}\,,\label{eq:VertexChConstraintt}\\
u=\left|\left(p_{2}-p_{3}\right)^{2}\right|=\left|\left(p_{1}-p_{4}\right)^{2}\right|\leq\left|\phi\right|^{2}\,\label{eq:VertexChConstraintu}
\end{align}
for a scattering amplitude mediated by a 4-vertex with momentum labeling
as defined in Fig.\,\ref{Gen4scat}. 
\noindent Specializing to the situation of Fig.\,\ref{fig:Massless,-massless-to hh u hh to hh}
and introducing $x\in\{p,P\}$ to cover both diagrams, we have 
\begin{align}
s= & \left|\left(R+S\right)^{2}\right|\leq\left|\phi\right|^{2}\,,\label{eq:SpezConstrs}\\
t= & \left|\left(x-R\right)^{2}\right|\leq\left|\phi\right|^{2}\,,\label{eq:SpezConstrt}\\
u= & \left|\left(x-S\right)^{2}\right|\leq\left|\phi\right|^{2}\,.\label{eq:SpezConstru}
\end{align}
Let us now show that it is impossible to satisfy each of these constraints
simultaneously. Appealing to Eqs.\,(\ref{eq:SpezConstrs}) through
(\ref{eq:SpezConstru}) and using $P^{2}=Q^{2}=R^{2}=S^{2}=m^{2}$,
$p^{2}=0$, one arrives at the following constraints for the case of two photons, two massive modes (left) and four massive modes (right):\\  

\noindent %
\begin{tabular}{c>{\centering}p{0.5\textwidth}}
\multicolumn{1}{>{\centering}p{0.5\textwidth}}{two photons, two massive modes: 

\begin{equation}
s=|2R_{0}S_{0}+2m^{2}-2\mathbf{R}\mathbf{S}|\leq|\phi|^{2}\,,\label{eq:ConstraintsA2-1-1}
\end{equation}

\begin{equation}
t=|-2p_{0}R_{0}+m^{2}+2\mathbf{p}\mathbf{R}|\leq|\phi|^{2}\,,\label{eq:ConstraintsA1-1-1}
\end{equation}

\begin{equation}
u=|-2p_{0}S_{0}+m^{2}+2\mathbf{p}\mathbf{S}|\leq|\phi|^{2}\,,\label{eq:ConstraintsA3-1-1}
\end{equation}
} & four massive modes:

\begin{equation}
s=|2R_{0}S_{0}+2m^{2}-2\mathbf{R}\mathbf{S}|\leq|\phi|^{2}\,,\,\,\,\,\label{eq:ConstraintsB2-1-1}
\end{equation}

\begin{equation}
t=|-2R_{0}P_{0}+2m^{2}+2\mathbf{R}\mathbf{P}|\leq|\phi|^{2}\,,\label{eq:ConstraintsB1-1}
\end{equation}

\begin{equation}
u=|-2S_{0}P_{0}+2m^{2}+2\mathbf{S}\mathbf{P}|\leq|\phi|^{2}\,.\label{eq:ConstraintsB3-1-1}
\end{equation}
\tabularnewline
\end{tabular}

\begin{figure}[h]
\begin{centering}
\includegraphics[width=1\textwidth]{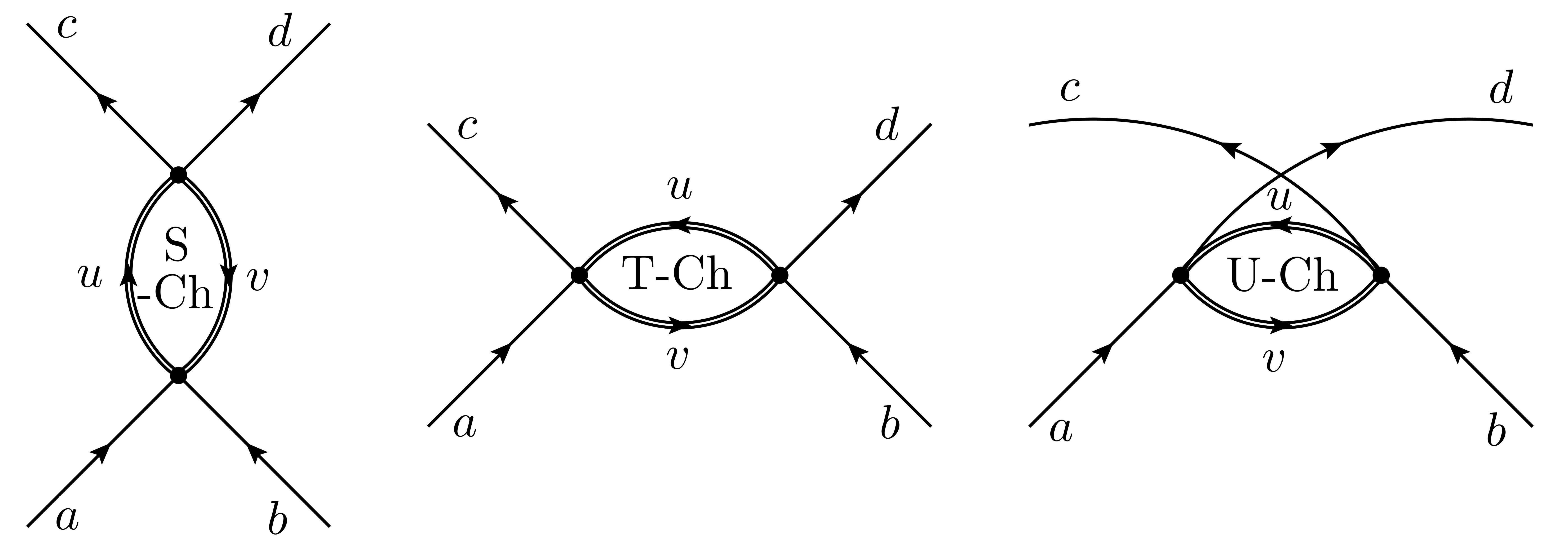} 
\par\end{centering}

\caption{\label{fig:All-3-Overall-Ch.} The overall scattering channels for the only 
admissible class of diagrams (two 4-vertices) in one-loop photon-photon scattering. Here
$a$,$b$,$c$, and $d$ are the four-momenta of the photons, and
$u$ and $v$ denote the loop four-momenta of the massive modes.}
\end{figure}

\noindent We have 
\begin{equation}
|2p_{0}R_{0}|>|2p_{0}(\sqrt{R_{0}^{2}-m^{2}})\cos\left(\measuredangle\left(\mathbf{p},\mathbf{R}\right)\right)|=|2\mathbf{p}\mathbf{R}|\,,\label{eq:4VertexAbschaetzung1}
\end{equation}
\begin{equation}
|2R_{0}S_{0}|>|2(\sqrt{R_{0}^{2}-m^{2}}\sqrt{S_{0}^{2}-m^{2}})\cos(\measuredangle\left(\mbox{\ensuremath{\mathbf{R}},\ensuremath{\mathbf{S}}}\right))|=|2\mathbf{R}\mathbf{S}|\,.\label{eq:4VertexAbschaetzung2}
\end{equation}
The following argument does not rely on a restriction of
the signs of the energies $p_{0}$, $R_{0}$, and $S_{0}$. From (\ref{eq:4VertexAbschaetzung2})
and the fact that $m^{2}>|\phi|^{2}$ it follows that the sign of
both $R_{0}$ and $S_{0}$ in (\ref{eq:ConstraintsA2-1-1}) need to
be different. Similar conclusions can be drawn for the other five cases
(\ref{eq:ConstraintsA1-1-1}) through (\ref{eq:ConstraintsB3-1-1}).
Again, using variable $x\in\{p,P\}$, we can summarize the situation
as 
\begin{align*}
s\leq|\phi|^{2}\Rightarrow\mbox{sgn}(R_{0}) & \,=-\mbox{sgn}(S_{0})\,,\\
t\leq|\phi|^{2}\Rightarrow\,\mbox{sgn}(x_{0}) & \,=\mbox{sgn}(R_{0})\,,\\
u\leq|\phi|^{2}\Rightarrow\,\mbox{sgn}(x_{0}) & \,=\mbox{sgn}(S_{0})\,.
\end{align*}
Therefore, one arrives at the following statement: 
\begin{align}
\mbox{sgn}(x_{0})=\mbox{sgn}(R_{0})=-\mbox{sgn}(S_{0})=-\mbox{sgn}(x_{0})\,.
\end{align}
As a consequence, the constraints (\ref{eq:ConstraintsA1-1-1}) through
(\ref{eq:ConstraintsA3-1-1}) or (\ref{eq:ConstraintsB2-1-1}) through
(\ref{eq:ConstraintsB3-1-1}) can not be satisfied simultaneously.
Thus, rather than imposing all constraints simultaneously at a given
4-vertex, a weighted superposition of allowed 2$\to$ 2 diagrams, satisfying only one momentum-transfer constraint for 
$s$, $t$, or $u$ at a time, is to be considered. As mentioned above, the
4-vertex is blind to permutations of legs, i.e.
it treats every scattering channel in the same way. For the case, where none of the constraints
(\ref{eq:ConstraintsA1-1-1}) through (\ref{eq:ConstraintsB3-1-1})
is trivial, this leads to the following prescription for the implementation
of a 4-vertex in the effective theory \cite{Hofmann2012}: 
\begin{align}
\Gamma_{\left[4\right]abcd}^{\alpha\beta\gamma\delta}=\frac{1}{3}\left(\left.\Gamma_{\left[4\right]abcd}^{\alpha\beta\gamma\delta}\right|_{s}+\left.\Gamma_{\left[4\right]abcd}^{\alpha\beta\gamma\delta}\right|_{t}+\left.\Gamma_{\left[4\right]abcd}^{\alpha\beta\gamma\delta}\right|_{u}\right)\,.\label{eq:Real4Vertex}
\end{align}
If, on the other hand, say, the $t$ channel is trivial (polarization
tensor of the massless mode on the one-loop level or figure-eight
two-loop contributions to the pressure \cite{SchwarzHofmannGiacosa2007,LudescherHofmann2009})
then the vertex acts as \cite{Hofmann2012}%
\footnote{It is easy to see that the integrand in this case is invariant under
$k\to-k$ ($k$ a loop four-momentum) which renders the contributions of
$s$ and $u$ channels equal. Thus it suffices to compute the $s$
channel contribution with weighting unity as it was done 
in \cite{SchwarzHofmannGiacosa2007,LudescherHofmann2009}.} 
\begin{align}
\Gamma_{\left[4\right]abcd}^{\alpha\beta\gamma\delta}=\frac{1}{2}\left(\left.\Gamma_{\left[4\right]abcd}^{\alpha\beta\gamma\delta}\right|_{s}+\left.\Gamma_{\left[4\right]abcd}^{\alpha\beta\gamma\delta}\right|_{u}\right)\,.\label{eq:Real4Vertex2}
\end{align}
This decomposition of a given 4-vertex into its scattering channels
is the reason for the large number of $3^{3}=27$ possible scattering
channel combinations in the three overall channels for one-loop photon-photon
scattering, compare with Fig.\,\ref{fig:All-3-Overall-Ch.} 

\subsection{Energy-flow constraints\label{sub:Energy-flow-constraints}}

Let us now investigate all possible constraints that can occur in
one-loop photon-photon scattering in view of energy flow. To do this,
it is advantageous to work with dimensionless quantities. We will
use two different normalizations. A tilde on top of a four-momentum
component $x$ marks normalizations with respect to the modulus of
the adjoint scalar field $\left|\phi\right|$, a hat denotes normalization
with respect to the temperature $T$, e.g. 
\begin{align}
\tilde{x}\equiv\frac{x}{\left|\phi\right|}=\frac{2e}{m}x=\frac{\lambda^{\frac{3}{2}}}{2\pi}\frac{x}{T}\equiv\frac{\lambda^{\frac{3}{2}}}{2\pi}\hat{x}\,,\label{eq:DimLos}
\end{align}
where the dimensionless temperature $\lambda$ is defined as 
\begin{align}
\lambda\equiv\frac{2\pi T}{\Lambda}\,.
\end{align}
Furthermore, the dimensionless mass $\tilde{m}\equiv\frac{m}{\left|\phi\right|}$
is given as 
\begin{align}
\tilde{m}= & 2e\,,\label{eq:mTilde}
\end{align}
where $e\geq\sqrt{8}\pi$ \cite{Hofmann2007}.

\noindent For a 4-vertex connecting two photons (four-momenta $\tilde{p}$,
$\tilde{q}$) with two massive modes (four-momenta $\tilde{R}$, $\tilde{S}$)
simple combinatorics allows the following possibilities of momentum transfer through the
vertex, redundant
by four-momentum conservation and constrained by the maximal resolution
$|\phi|$ of the effective theory: 
\begin{align}
\left|\left(\tilde{R}\pm\tilde{S}\right)^{2}\right| & \leq1\label{eq:CombisAngang}\\
\left|\left(\tilde{p}\pm\tilde{R}\right)^{2}\right| & \leq1\label{eq:Combis3}\\
\left|\left(\tilde{p}\pm\tilde{S}\right)^{2}\right| & \leq1\label{eq:Combis3p}\\
\left|\left(\tilde{q}\pm\tilde{R}\right)^{2}\right| & \leq1\label{eq:Combis5}\\
\left|\left(\tilde{q}\pm\tilde{S}\right)^{2}\right| & \leq1\label{eq:Combis5p}\\
\left|\left(\tilde{p}\pm\tilde{q}\right)^{2}\right| & \leq1\,.\label{eq:CombisEnde-1}
\end{align}
Both, photons and massive modes are on shell: 
\begin{align}
\tilde{R}_{0} & =\pm\sqrt{\left|\mathbf{\mathbf{\tilde{R}}}\right|^{2}+\tilde{m}^{2}}\,,\label{inionshee}\\
\tilde{S}_{0} & =\pm\sqrt{\left|\mathbf{\mathbf{\tilde{S}}}\right|^{2}+\tilde{m}^{2}}\,,\\
\tilde{p}^{2} & =0\,,\\
\tilde{q}^{2} & =0\,.\label{finonshee}
\end{align}
>From Eqs.\,(\ref{eq:mTilde}) and (\ref{inionshee}) through (\ref{finonshee})
it follows that certain sign combinations of $R_{0}$ and $S_{0}$
are forbidden. Let us now classify those excluded cases. We consider 
\begin{align}
1\geq & \left|\left(\tilde{R}\pm\tilde{S}\right)^{2}\right|=\left|2\tilde{m}^{2}\pm2\tilde{R}_{0}\tilde{S}_{0}\mp2\mathbf{\tilde{R}}\mathbf{\tilde{S}}\right|\nonumber \\
= & \left|2\tilde{m}^{2}\pm 2\left(\pm\sqrt{\left|\mathbf{\tilde{R}}\right|^{2}+\tilde{m}^{2}}\right)\left(\pm\sqrt{\left|\mathbf{\tilde{S}}\right|^{2}+
\tilde{m}^{2}}\right)\mp2\mathbf{\tilde{R}}\mathbf{\tilde{S}}\right| \label{massivemassivesigns}
\end{align}
and 
\begin{align}
1\geq & \left|\left(\tilde{p}\pm\tilde{R}\right)^{2}\right|=\left|\tilde{m}^{2}\pm2\tilde{p}_{0}\tilde{R}_{0}\mp2\mathbf{\tilde{p}}\mathbf{\tilde{R}}\right|\nonumber \\
= & \left|\tilde{m}^{2}\pm2\tilde{p}_{0}\left(\pm\sqrt{\left|\mathbf{\tilde{R}}\right|^{2}+\tilde{m}^{2}}\right)\mp2\mathbf{\tilde{p}}\mathbf{\tilde{R}}\right|\,.\label{eq:photonmassivesigns}
\end{align}
In the following discussion cases \eqref{eq:CombisAngang} are treated
in terms of (\ref{massivemassivesigns}) while cases (\ref{eq:Combis3})
through (\ref{eq:Combis5p}) are covered%
\footnote{We do not distinguish $R$ and $S$ or $p$ and $q$. For the argument
it is only important that the former are associated with massive modes
and the latter with photons.%
} by (\ref{eq:photonmassivesigns}). To proceed, note that 
\begin{align}
\left|\left(\pm\sqrt{\left|\mathbf{\tilde{R}}\right|^{2}+\tilde{m}^{2}}\right)\left(\pm\sqrt{\left|\mathbf{\tilde{S}}\right|^{2}+\tilde{m}^{2}}\right)\right|\mp\mathbf{\tilde{R}}\mathbf{\tilde{S}}\geq\tilde{m}^{2}\label{modmassivemixed}
\end{align}
and
\noindent 
\begin{align}
\left|p_{0}\left(\pm\sqrt{\left|\mathbf{\tilde{R}}\right|^{2}+\tilde{m}^{2}}\right)\right|\mp\mathbf{\tilde{p}}\mathbf{\tilde{R}}\geq0\label{modmasslessmixed}
\end{align}
\noindent 
Inequality \eqref{modmassivemixed} is true because of the Cauchy-Schwarz
inequality applied to the two vectors $\left(\left|\mathbf{\tilde{R}}\right|,\tilde{m}\right)$
and $\left(\left|\mathbf{\tilde{S}}\right|,\tilde{m}\right)$ with
the canonical scalar product of $\mathbb{R}^{2}$ and the fact that
$\left|\cos\measuredangle\mathbf{\tilde{R}}\mathbf{\tilde{S}}\right|\leq1$. 
Inequality \eqref{modmasslessmixed} is selfevident. 
>From \eqref{modmassivemixed}, $m^{2}>1$ (see \eqref{eq:mTilde}),
and \eqref{massivemassivesigns} it follows that
\begin{align*}
\mbox{sgn}\left(\tilde{R}_{0}\right)=\pm \mbox{sgn}\left(\tilde{S}_{0}\right)
\end{align*}
are forbidden for cases \eqref{eq:CombisAngang}, respectively. 
Because of \eqref{modmasslessmixed}, $\tilde{m}^{2}>1$, and \eqref{eq:photonmassivesigns}
it is clear that
\begin{align*}
\mbox{sgn}\left(\tilde{R}_{0}\right)=\pm 1
\end{align*}
are forbidden for cases \eqref{eq:Combis3}, respectively (and for cases \eqref{eq:Combis5}, respectively).
In addition, the respective cases \eqref{eq:Combis3p} and \eqref{eq:Combis5p}
($\tilde{R}$ replaced by $\tilde{S}$) are also excluded. Obviously, 
no implication for the signs of $\tilde{R}_{\text{0}}$ or $\tilde{S}_{\text{0}}$
arises from \eqref{eq:CombisEnde-1}. 

\subsection{Counting of constraints}

Because a 4-vertex is constrained by one of the constraints (\ref{eq:VertexChConstraints}), (\ref{eq:VertexChConstraintt}), or 
(\ref{eq:VertexChConstraintu}) only the estimate of the ratio of the number $\tilde{K}$ of independent, radial loop variables 
versus the number $K$ of constraints on them obtained in Eq.\,(5.101) of  
\cite{Hofmann2012} or in Eq.\,(16) of \cite{Hofmann2006} modifies as
\eqb\label{modconstraintscounting}
\frac{\tilde{K}}{K}\le \frac45\left(1+\frac{1}{V_4}\right)
\eqe
for the case of a planar diagram containing only $V_4$ many 4-vertices. This generalises to 
\eqb\label{modconstraintscountingNP}
\frac{\tilde{K}}{K}\le \frac45\left(1+\frac{1}{V_4}(1-2g)\right)
\eqe
if the diagram exhibits a nontrivial genus $g$. For $g=0$ one has $\frac{\tilde{K}}{K}\le 1$ 
provided that $V_4\ge 4$.

\section{Photon-photon scattering\label{Pps}}

Let us now assess the options for one-loop photon-photon scattering.
To do this, we first apply the results of Sec.\,\ref{sub:Energy-flow-constraints}
to each of the overall scattering channels depicted in Fig. \ref{fig:All-3-Overall-Ch.}.
This excludes a vast majority of a priori thinkable energy-flow combinations.
In the overall S-channel scattering channel all
combinations for the two 4-vertices can be excluded by analytical
arguments. For the overall T-channel (U-channel as well) the analytical treatment leaves four
combinations possible.
These are investigated by numerical analysis
based on Monte-Carlo simulations.


At a given vertex, we use the following notation to keep track of
excluded combinations of the two loop energies $u_{0}$ and $v_{0}$.
Every cell in a $2\times2$ table accounts for one possible combinations
of energy flow: 

\noindent \begin{center}
\begin{tabular}{|c|c|}
\hline 
$\tilde{u}_{0}>0;\:\tilde{v}_{0}>0$ & $\tilde{u}_{0}>0;\:\tilde{v}_{0}<0$\tabularnewline
\hline 
$\tilde{u}_{0}<0;\:\tilde{v}_{0}>0$ & $\tilde{u}_{0}<0;\:\tilde{v}_{0}<0$\tabularnewline
\hline 
\end{tabular}.
\par\end{center}
\noindent An ``X'' in a given cell signals that the corresponding
combination is forbidden.

\subsection{Energy flow: Overall S channel\label{sub:Energy-flow:-Overall}}

\noindent 
\begin{figure}[h]
\begin{centering}
\includegraphics[width=0.3\textwidth]{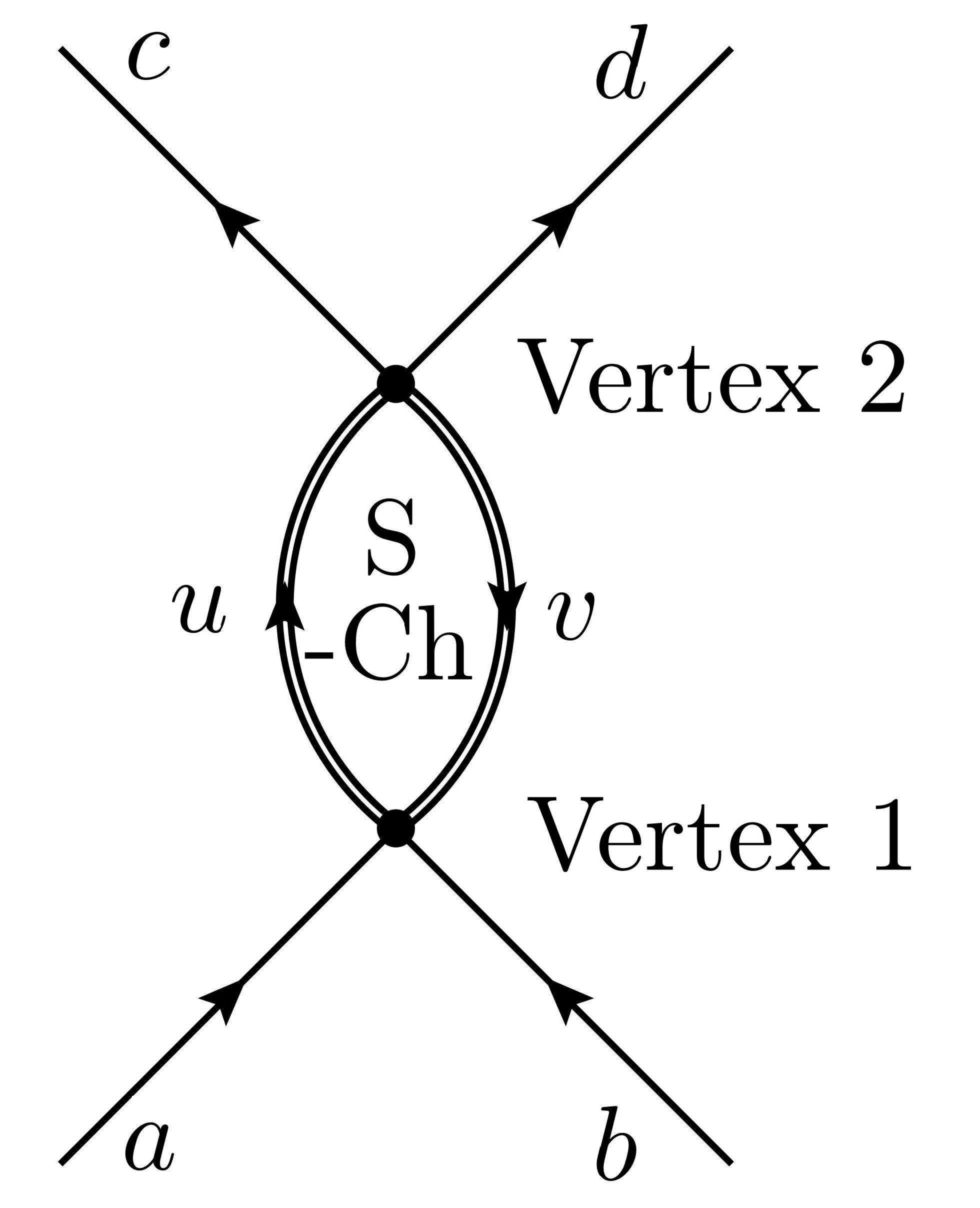}
\par\end{centering}

\caption{\label{fig:The-overall-S-channel}The overall S-channel.}
\end{figure}

\noindent The lower vertex in Fig. \ref{fig:The-overall-S-channel}
is denoted by number 1, the upper one by number 2. At a given vertex
the constraint on a scattering channel can be expressed in a twofold
way because of total four-momentum conservation across the vertex.
For example, the constraint at the first vertex $\left|\left(\tilde{a}+\tilde{b}\right)^{2}\right|\leq1$
can be rewritten as $\left|\left(\tilde{u}-\tilde{v}\right)^{2}\right|\leq1$.
To exclude all sign combinations of the energies in the loop momenta
one needs to look at either form and combine both statements
\footnote{The two $2\times2$ tables, obtained from each of the forms expressing
the vertex constraint, are put on top of one another.}.

\noindent Let us now visualize the constraints at vertex 1 in terms
of their tables:
\begin{center}
$\textnormal{s-ch:}\;\;1\geq\left|\left(\tilde{a}+\tilde{b}\right)^{2}\right|=\left|\left(\tilde{u}-\tilde{v}\right)^{2}\right|\rightarrow\:$%
\begin{tabular}{|c|c|}
\hline 
 & X\tabularnewline
\hline 
X & \tabularnewline
\hline 
\end{tabular}$\,,$
\par\end{center}

\begin{center}
$\textnormal{t-ch:}\;\;1\geq\left|\left(\tilde{a}-\tilde{u}\right)^{2}\right|=\left|\left(\tilde{b}+\tilde{v}\right)^{2}\right|\rightarrow\:$%
\begin{tabular}{|c|c|}
\hline 
X & \tabularnewline
\hline 
X & X\tabularnewline
\hline 
\end{tabular}$\,,$
\par\end{center}

\begin{center}
$\textnormal{u-ch:}\;\;1\geq\left|\left(\tilde{a}+\tilde{v}\right)^{2}\right|=\left|\left(\tilde{b}-\tilde{u}\right)^{2}\right|\rightarrow\:$%
\begin{tabular}{|c|c|}
\hline 
X & \tabularnewline
\hline 
X & X\tabularnewline
\hline 
\end{tabular}$\,.$
\par\end{center}

\noindent For vertex 2 we obtain:

\begin{center}
$\textnormal{s-ch:}\;\;1\geq\left|\left(\tilde{u}-\tilde{v}\right)^{2}\right|=\left|\left(\tilde{c}+\tilde{d}\right)^{2}\right|\rightarrow\:$%
\begin{tabular}{|c|c|}
\hline 
 & X\tabularnewline
\hline 
X & \tabularnewline
\hline 
\end{tabular}$\,,$
\par\end{center}

\begin{center}
$\textnormal{t-ch:}\;\;1\geq\left|\left(\tilde{u}-\tilde{c}\right)^{2}\right|=\left|\left(\tilde{v}+\tilde{d}\right)^{2}\right|\rightarrow\:$%
\begin{tabular}{|c|c|}
\hline 
X & \tabularnewline
\hline 
X & X\tabularnewline
\hline 
\end{tabular}$\,,$
\par\end{center}

\begin{center}
$\textnormal{u-ch:}\;\;1\geq\left|\left(\tilde{u}-\tilde{d}\right)^{2}\right|=\left|\left(\tilde{v}+\tilde{c}\right)^{2}\right|\rightarrow\:$%
\begin{tabular}{|c|c|}
\hline 
X & \tabularnewline
\hline 
X & X\tabularnewline
\hline 
\end{tabular}$\,.$
\par\end{center}

\noindent In the diagram for the overall scattering channel (Fig.
\ref{fig:The-overall-S-channel}) the constraints on the two vertices
have to be satisfied together in order to contribute: If one combination
of energy flow and scattering channel is forbidden by
one constraint then it is not allowed at all. Thus it is suggested
to use a $3\times3$ table for visualization of excluded scattering-channel
and energy-flow combinations within a given overall channel. Each
of the cells of this $3\times3$ table, corresponding to a certain
combination of scattering channels at vertex 1 and vertex 2, is obtained
by putting the two respective $2\times2$ tables on top of one another.
For the overall S-channel we thus are left with Tab. \ref{tab:TableOverallSXX}.

\noindent 
\begin{table}[H]
\centering{}%
\begin{tabular}{|c|c|c|c|c|c|c|c|c|c|}
\cline{1-1} \cline{3-4} \cline{6-7} \cline{9-10} 
\backslashbox{Vertex 2}{Vertex 1} &  & \multicolumn{2}{c|}{s-ch.} &  & \multicolumn{2}{c|}{t-ch.} &  & \multicolumn{2}{c|}{u-ch.}\tabularnewline
\cline{1-1} \cline{3-4} \cline{6-7} \cline{9-10} 
\multicolumn{1}{c}{} & \multicolumn{1}{c}{} & \multicolumn{1}{c}{} & \multicolumn{1}{c}{} & \multicolumn{1}{c}{} & \multicolumn{1}{c}{} & \multicolumn{1}{c}{} & \multicolumn{1}{c}{} & \multicolumn{1}{c}{} & \multicolumn{1}{c}{}\tabularnewline
\cline{1-1} \cline{3-4} \cline{6-7} \cline{9-10} 
\multirow{2}{*}{s- ch.} &  &  & X &  & X & X &  & X & X\tabularnewline
\cline{3-4} \cline{6-7} \cline{9-10} 
 &  & X &  &  & X & X &  &  X & X\tabularnewline
\cline{1-1} \cline{3-4} \cline{6-7} \cline{9-10} 
\multicolumn{1}{c}{} & \multicolumn{1}{c}{} & \multicolumn{1}{c}{} & \multicolumn{1}{c}{} & \multicolumn{1}{c}{} & \multicolumn{1}{c}{} & \multicolumn{1}{c}{} & \multicolumn{1}{c}{} & \multicolumn{1}{c}{} & \multicolumn{1}{c}{}\tabularnewline
\cline{1-1} \cline{3-4} \cline{6-7} \cline{9-10} 
\multirow{2}{*}{t-ch.} &  & X & X &  & X &  &  & X & \tabularnewline
\cline{3-4} \cline{6-7} \cline{9-10} 
 &  & X & X &  & X & X &  & X & X\tabularnewline
\cline{1-1} \cline{3-4} \cline{6-7} \cline{9-10} 
\multicolumn{1}{c}{} & \multicolumn{1}{c}{} & \multicolumn{1}{c}{} & \multicolumn{1}{c}{} & \multicolumn{1}{c}{} & \multicolumn{1}{c}{} & \multicolumn{1}{c}{} & \multicolumn{1}{c}{} & \multicolumn{1}{c}{} & \multicolumn{1}{c}{}\tabularnewline
\cline{1-1} \cline{3-4} \cline{6-7} \cline{9-10} 
\multirow{2}{*}{u-ch.} &  & X & X &  & X &  &  & X & \tabularnewline
\cline{3-4} \cline{6-7} \cline{9-10} 
 &  & X & X &  & X & X &  & X & X\tabularnewline
\cline{1-1} \cline{3-4} \cline{6-7} \cline{9-10} 
\end{tabular}\caption{\label{tab:TableOverallSXX}Forbidden combinations of energy flow
(marked with a X) in all scattering channel combinations of vertex
1 and vertex 2 in the overall S-channel.}
\end{table}

\noindent Therefore, only six possible combinations remain. Two of them can be excluded analytically as we will show in the
next section.

\subsection{Exclusion of configuration Sss\label{sub:Exclusion of Sss}}

\noindent For the configuration Sss two combinations of energy flow 
are allowed by Tab. \ref{tab:TableOverallSXX}. To exclude them, we 
proceed by a more detailed analysis of the general situation
expressed by Eqs.\,\eqref{eq:CombisEnde-1} and \eqref{inionshee} to \eqref{finonshee}. 
For our specific cases one has
\noindent 
\begin{align}
 & \left|\left(\tilde{a}+\tilde{b}\right)^{2}\right|=\left|2\tilde{a}\tilde{b}\right|=2\left|\tilde{a}_{0}\tilde{b}_{0}(1-\cos\left(\measuredangle\mathbf{\tilde{a}}\mathbf{\tilde{b}}\right))\right|\leq1\nonumber \\
\Rightarrow & -\frac{1}{2}\leq\tilde{a}_{0}\tilde{b}_{0}(1-\cos\left(\measuredangle\mathbf{\tilde{a}}\mathbf{\tilde{b}}\right))\leq\frac{1}{2}\nonumber \\
\Rightarrow & -\frac{1}{2\tilde{a}_{0}\tilde{b}_{0}}\leq(1-\cos\left(\measuredangle\mathbf{\tilde{a}}\mathbf{\tilde{b}}\right))\leq\frac{1}{2\tilde{a}_{0}\tilde{b}_{0}}\,.\label{eq:SssMomConstr}
\end{align}
The first inequality $-\frac{1}{2\tilde{a}_{0}\tilde{b}_{0}}\leq(1-\cos\measuredangle\mathbf{\tilde{a}}\mathbf{\tilde{b}})$
is always satisfied because $\left|\cos\measuredangle\mathbf{\tilde{a}}\mathbf{\tilde{b}}\right|\leq1$ and because $a_0,b_0>0$.
We also have 
\begin{align}
\tilde{u}^{2}=\tilde{m}^{2}\,,\\
\tilde{v}^{^{2}}=\tilde{m}^{2}\,.
\end{align}
\noindent In addition, four-momentum conservation holds.
Thus the a priori eight independent entries of the two four-momenta $\tilde{u}$ and $\tilde{v}$ are
reduced to four. The on-shellness of each mode leads to an additional
reduction from four to two: $\left|\mathbf{\tilde{u}}\right|$ is
determined by $\tilde{a}$, $\tilde{b}$, and the orientation
$\mathbf{e_{u}}=\frac{\mathbf{u}}{\left|\mathbf{u}\right|}=\frac{\tilde{\mathbf{u}}}{\left|\tilde{\mathbf{u}}\right|}$
(two angles). One has
\noindent 
\begin{align}
0= & \tilde{v}^{2}-\tilde{m}^{2}=\left(\tilde{u}-\tilde{a}-\tilde{b}\right)^{2}-\tilde{m}^{2}=-2\tilde{u}\left(\tilde{a}+\tilde{b}\right)+2\tilde{a}\tilde{b}\nonumber \\
= & -2\left(\pm\sqrt{\left|\mathbf{\tilde{u}}\right|^{2}+\tilde{m}^{2}}\right)\left(\tilde{a}_{0}+\tilde{b}_{0}\right)+2\mathbf{\tilde{u}}\left(\mathbf{\tilde{a}}+\mathbf{\tilde{b}}\right)+2\tilde{a}\tilde{b}\nonumber \\
\Rightarrow\phantom{0=} & \left(\pm\sqrt{\left|\mathbf{\tilde{u}}\right|^{2}+\tilde{m}^{2}}\right)\left(\tilde{a}_{0}+\tilde{b}_{0}\right)=\left|\mathbf{\tilde{u}}\right|\left(\mathbf{\tilde{a}}+\mathbf{\tilde{b}}\right)\mathbf{e_{u}}+\tilde{a}\tilde{b}\nonumber \\
\Rightarrow\phantom{0=} & \left(\left|\mathbf{\tilde{u}}\right|^{2}+\tilde{m}^{2}\right)\left(\tilde{a}_{0}+\tilde{b}_{0}\right)^{2}\nonumber \\
= & \left|\mathbf{\tilde{u}}\right|^{2}\left(\left(\mathbf{\tilde{a}}+\mathbf{\tilde{b}}\right)\mathbf{e_{u}}\right)^{2}+\left(\tilde{a}\tilde{b}\right)^{2}+2\left|\mathbf{\tilde{u}}\right|\left(\tilde{a}\tilde{b}\right)\left(\mathbf{\tilde{a}}+\mathbf{\tilde{b}}\right)\mathbf{e_{u}}\nonumber \\
\Rightarrow0= & \left|\mathbf{\tilde{u}}\right|^{2}\left(\left(\tilde{a}_{0}+\tilde{b}_{0}\right)^{2}-\left(\left(\mathbf{\tilde{a}}+\mathbf{\tilde{b}}\right)\mathbf{e_{u}}\right)^{2}\right)+\left|\mathbf{\tilde{u}}\right|\left(-2\left(\tilde{a}\tilde{b}\right)\left(\mathbf{\tilde{a}}+\mathbf{\tilde{b}}\right)\mathbf{e_{u}}\vphantom{\left(\left(\vec{x}+\vec{x}\right)\vec{\tilde{x}}_{n}\right)^{2}}\right)\nonumber \\
 & +\left(\tilde{m}^{2}\left(\tilde{a}_{0}+\tilde{b}_{0}\right)^{2}-\left(\tilde{a}\tilde{b}\right)^{2}\right)\nonumber \\
\Rightarrow\left|\mathbf{\tilde{u}}\right|_{1/2}= & \frac{\left(\tilde{a}\tilde{b}\right)\left(\mathbf{\tilde{a}}+\mathbf{\tilde{b}}\right)\mathbf{e_{u}}}{\left(\tilde{a}_{0}+\tilde{b}_{0}\right)^{2}-\left(\left(\mathbf{\tilde{a}}+\mathbf{\tilde{b}}\right)\mathbf{e_{u}}\right)^{2}}\nonumber \\
 & \pm\frac{\left(\tilde{a}_{0}+\tilde{b}_{0}\right)\sqrt{-\tilde{m}^{2}\left(\tilde{a}_{0}+
\tilde{b}_{0}\right)^{2}+\left(\tilde{a}\tilde{b}\right)^{2}+\tilde{m}^{2}
\left(\left(\mathbf{\tilde{a}}+\mathbf{\tilde{b}}\right)\mathbf{e_{u}}\right)^{2}}}{\left(\tilde{a}_{0}+
\tilde{b}_{0}\right)^{2}-\left(\left(\mathbf{\tilde{a}}+\mathbf{\tilde{b}}\right)\mathbf{e_{u}}\right)^{2}}\,.\label{eq:iniLsg1/2}
\end{align}
In the computation of the amplitude only an integration over the orientation
$\mathbf{e_{u}}$ remains when it comes to loop integration. The solutions $\left|\mathbf{\tilde{u}}\right|_{1/2}$ in
Eq. \eqref{eq:iniLsg1/2} must be real and positive. This implies
that valid configurations satisfy the following inequality (argument
of square root in Eq.\,(\ref{eq:iniLsg1/2}) must be positive):

\noindent 
\begin{align}
 & -\tilde{m}^{2}\left(\tilde{a}_{0}+\tilde{b}_{0}\right)^{2}+\left(\tilde{a}\tilde{b}\right)^{2}+
\tilde{m}^{2}\left(\left(\mathbf{\tilde{a}}+\mathbf{\tilde{b}}\right)\mathbf{e_{u}}\right)^{2}\geq0\nonumber \\
\Rightarrow\phantom{0=} & -\tilde{m}^{2}\left(\tilde{a}_{0}^{2}+\tilde{b}_{0}^{2}+2\tilde{a}_{\text{0}}\tilde{b}_{\text{0}}\right)+
\tilde{a}_{0}^{2}\tilde{b}_{0}^{2}\left(1-\cos\left(\measuredangle\mathbf{a}\mathbf{b}\right)\right)^{2}\nonumber \\
 & +\tilde{m}^{2}\left(\tilde{a}_{0}^{2}+\tilde{b}_{0}^{2}+2\tilde{a}_{\text{0}}\tilde{b}_{\text{0}}
\cos\left(\measuredangle\mathbf{a}\mathbf{b}\right)\right)\left(\mathbf{e_{u}}\mathbf{e_{a+b}}\right)^2\geq0\,.
\end{align}
The values of $\mathbf{e_{u}}\mathbf{e_{a+b}}$, at which $\measuredangle\mathbf{a}\mathbf{b}$ 
is least constrained, are $\mathbf{e_{u}}\mathbf{e_{a+b}}=\pm 1$.
For these values the inequality reads as
\begin{align}
\tilde{a}_{0}^{2}\tilde{b}_{0}^{2}\left(1-\cos\left(\measuredangle\mathbf{a}\mathbf{b}\right)\right)^{2}-2\tilde{m}^{2}\tilde{a}_{0}\tilde{b}_{0}\left(1-\cos\left(\measuredangle\mathbf{a}\mathbf{b}\right)\right)\geq0\,.\label{eq:CosIneq}
\end{align}
To proceed, we need the roots $\left(1-\cos\left(\measuredangle\mathbf{a}\mathbf{b}\right)\right)_{1/2}$
of the equation
\begin{align}
\tilde{a}_{0}^{2}\tilde{b}_{0}^{2}\left(1-\cos\left(\measuredangle\mathbf{a}\mathbf{b}\right)\right)^{2}-2\tilde{m}^{2}\tilde{a}_{0}\tilde{b}_{0}\left(1-\cos\left(\measuredangle\mathbf{a}\mathbf{b}\right)\right)=0\,.
\end{align}
These roots are:
\begin{align}
\left(1-\cos\left(\measuredangle\mathbf{a}\mathbf{b}\right)\right)_{1} & =0\,,\\
\left(1-\cos\left(\measuredangle\mathbf{a}\mathbf{b}\right)\right)_{2} & =\frac{2\tilde{m}^{2}}{\tilde{a}_{0}\tilde{b}_{0}}\,.
\end{align}
Inequality \eqref{eq:CosIneq} either is solved by values of $\left(1-\cos\left(\measuredangle\mathbf{a}\mathbf{b}\right)\right)$
that are smaller than the first solution or larger than the second
one because with respect to the variable $\left(1-\cos\left(\measuredangle\mathbf{a}\mathbf{b}\right)\right)$
the left hand side of \eqref{eq:CosIneq} is a parabola with positive
curvature. $\left(1-\cos\left(\measuredangle\mathbf{a}\mathbf{b}\right)\right)_{1}<0$
can not be satisfied, and therefore every valid $\left(1-\cos\left(\measuredangle\mathbf{a}\mathbf{b}\right)\right)$
has to be larger than $\frac{2\tilde{m}^{2}}{\tilde{a}_{0}\tilde{b}_{0}}$:

\begin{equation}
\left(1-\cos\left(\measuredangle\mathbf{a}\mathbf{b}\right)\right)\geq\frac{2\tilde{m}^{2}}{\tilde{a}_{0}\tilde{b}_{0}}\,.\label{eq:1MinCosIneq2}
\end{equation}
\noindent We can now compare the two requirements for $\left(1-\cos\left(\measuredangle\mathbf{a}\mathbf{b}\right)\right)$.
\begin{itemize}
\item From on-shellness and momentum conservation at vertex 1 (Eq. \eqref{eq:1MinCosIneq2}):\\
\begin{align*}
 & \left(1-\cos\left(\measuredangle\mathbf{a}\mathbf{b}\right)\right)\geq\frac{2\tilde{m}^{2}}{\tilde{a}_{0}\tilde{b}_{0}}\,.
\end{align*}

\item From the s-channel momentum transfer constraint at vertex 1 (Eq. \eqref{eq:SssMomConstr}):
\begin{align*}
 & \,\,\,\,(1-\cos\left(\measuredangle\mathbf{a}\mathbf{b}\right))\leq\frac{1}{2\tilde{a}_{0}\tilde{b}_{0}}\,.
\end{align*}
\end{itemize}
\noindent The upper bound is smaller than the lower bound because
$\tilde{m}=2e\geq2\sqrt{8}\pi>1$, and so we conclude that there are
no configurations satisfying the momentum transfer constraint, on-shellness of the massive modes, 
and momentum conservation simultaneously. Thus the combination of s
and s from the two 4-vertices to the overall S-channel is excluded in photon-photon scattering, and we are left 
with four unexcluded configurations.

\subsection{Energy flow: Overall T- and U-channels}

\noindent Similar arguments as used for the overall S-channel
can be applied to the overall T- and U-channels. The following considerations do not depend on the specific
values of the external momenta but only on the fact that they are
on shell. The overall U-channel thus is constrained in the same way as the overall
T-channel. (One just interchanges the external momenta $c$ and $d$, see
Fig. \eqref{fig:All-3-Overall-Ch.}). Let us now put forward the arguments for 
the T-channel.

\noindent 
\begin{figure}[h]
\begin{centering}
\includegraphics[width=0.5\textwidth]{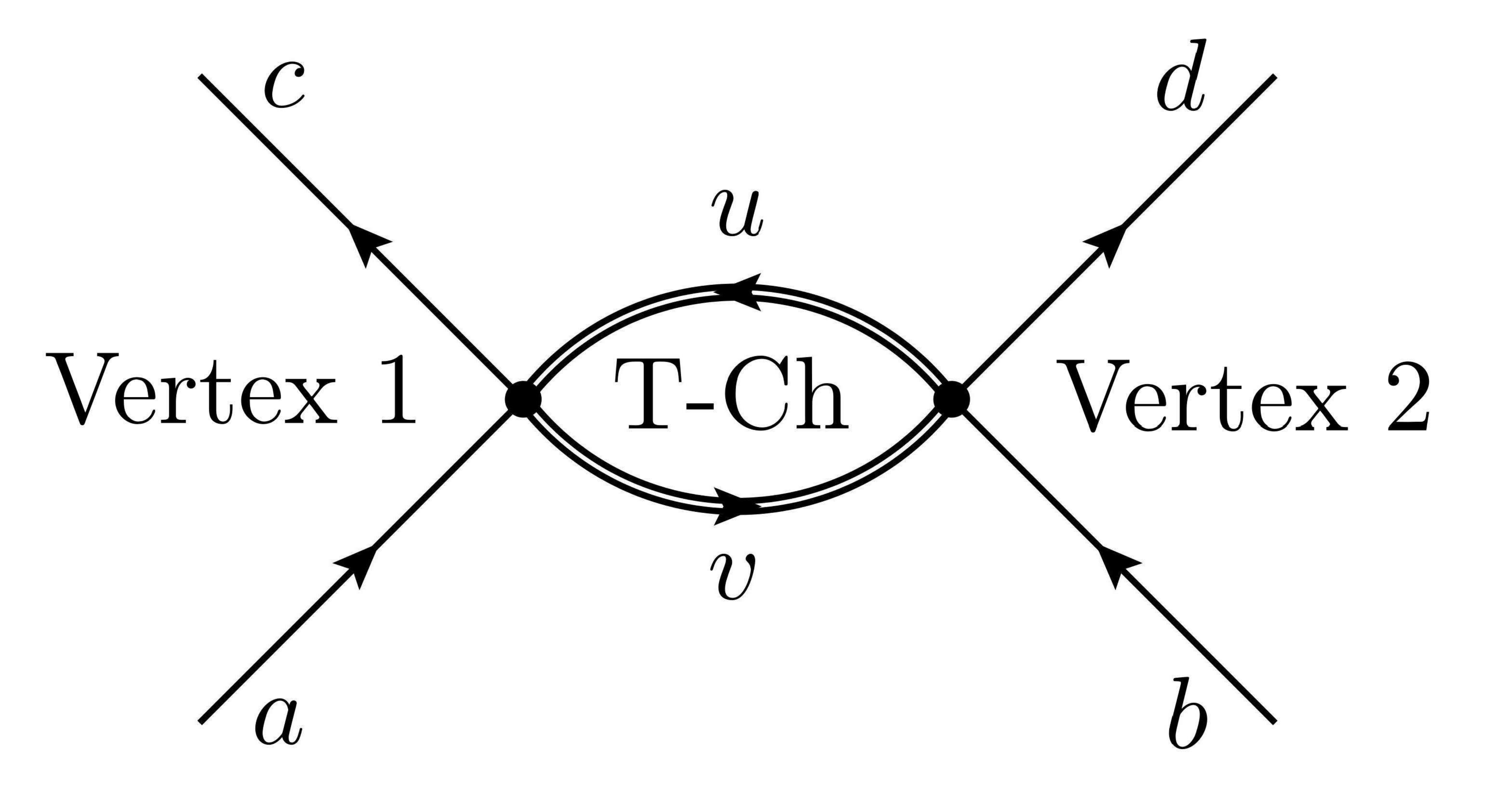}
\par\end{centering}

\caption{\label{fig:Overall-T-channel}Overall T-channel. The vertex on the left side is labeled by number 1, the one on the
right side by number 2.}
\end{figure}

\noindent The exclusion tables for the constraints at vertex 1 can
 be obtained by referring to Sec.\,\ref{sub:Energy-flow-constraints}:

\begin{center}
$\textnormal{s-ch:}\;\;1\geq\left|\left(\tilde{a}-\tilde{v}\right)^{2}\right|=\left|\left(\tilde{c}-\tilde{u}\right)^{2}\right|\rightarrow\:$%
\begin{tabular}{|c|c|}
\hline 
 & X\tabularnewline
\hline 
X & X\tabularnewline
\hline 
\end{tabular}$\,,$
\par\end{center}

\begin{center}
$\textnormal{t-ch:}\;\;1\geq\left|\left(\tilde{a}-\tilde{c}\right)^{2}\right|=\left|\left(\tilde{v}-\tilde{u}\right)^{2}\right|\rightarrow\:$%
\begin{tabular}{|c|c|}
\hline 
 & X\tabularnewline
\hline 
 X & \tabularnewline
\hline 
\end{tabular}$\,,$
\par\end{center}

\begin{center}
$\textnormal{u-ch:}\;\;1\geq\left|\left(\tilde{a}+\tilde{u}\right)^{2}\right|=\left|\left(\tilde{v}+\tilde{c}\right)^{2}\right|\rightarrow\:$%
\begin{tabular}{|c|c|}
\hline 
X & X\tabularnewline
\hline 
X & \tabularnewline
\hline 
\end{tabular}$\,.$
\par\end{center}

\noindent The constraints at vertex 2 read:

\begin{center}
$\textnormal{s-ch:}\;\;1\geq\left|\left(\tilde{v}+\tilde{b}\right)^{2}\right|=\left|\left(\tilde{u}+\tilde{d}\right)^{2}\right|\rightarrow\:$%
\begin{tabular}{|c|c|}
\hline 
X & X\tabularnewline
\hline 
 X& \tabularnewline
\hline 
\end{tabular}$\,,$
\par\end{center}

\begin{center}
$\textnormal{t-ch:}\;\;1\geq\left|\left(\tilde{v}-\tilde{u}\right)^{2}\right|=\left|\left(\tilde{b}-\tilde{d}\right)^{2}\right|\rightarrow\:$%
\begin{tabular}{|c|c|}
\hline 
 & X\tabularnewline
\hline 
X & \tabularnewline
\hline 
\end{tabular}$\,,$
\par\end{center}

\begin{center}
$\textnormal{u-ch:}\;\;1\geq\left|\left(\tilde{v}-\tilde{d}\right)^{2}\right|=\left|\left(\tilde{b}-\tilde{u}\right)^{2}\right|\rightarrow\:$%
\begin{tabular}{|c|c|}
\hline 
 & X\tabularnewline
\hline 
X & X\tabularnewline
\hline 
\end{tabular}$\,.$
\par\end{center}

\noindent In analogy to Sec.\,\ref{sub:Energy-flow:-Overall}, we superimpose all combinations of the tables for the two vertices to obtain the 
result shown in Tab. \ref{tab:By-momentum-transfer}.

\noindent 
\begin{table}[H]
\centering{}%
\begin{tabular}{|c|c|c|c|c|c|c|c|c|c|}
\cline{1-1} \cline{3-4} \cline{6-7} \cline{9-10} 
\backslashbox{Vertex 2}{Vertex 1} &  & \multicolumn{2}{c|}{s-ch.} &  & \multicolumn{2}{c|}{t-ch.} &  & \multicolumn{2}{c|}{u-ch.}\tabularnewline
\cline{1-1} \cline{3-4} \cline{6-7} \cline{9-10} 
\multicolumn{1}{c}{} & \multicolumn{1}{c}{} & \multicolumn{1}{c}{} & \multicolumn{1}{c}{} & \multicolumn{1}{c}{} & \multicolumn{1}{c}{} & \multicolumn{1}{c}{} & \multicolumn{1}{c}{} & \multicolumn{1}{c}{} & \multicolumn{1}{c}{}\tabularnewline
\cline{1-1} \cline{3-4} \cline{6-7} \cline{9-10} 
\multirow{2}{*}{s- ch.} &  & X & X &  & X & X &  & X & X\tabularnewline
\cline{3-4} \cline{6-7} \cline{9-10} 
 &  & X & X &  & X &  &  & X & \tabularnewline
\cline{1-1} \cline{3-4} \cline{6-7} \cline{9-10} 
\multicolumn{1}{c}{} & \multicolumn{1}{c}{} & \multicolumn{1}{c}{} & \multicolumn{1}{c}{} & \multicolumn{1}{c}{} & \multicolumn{1}{c}{} & \multicolumn{1}{c}{} & \multicolumn{1}{c}{} & \multicolumn{1}{c}{} & \multicolumn{1}{c}{}\tabularnewline
\cline{1-1} \cline{3-4} \cline{6-7} \cline{9-10} 
\multirow{2}{*}{t-ch.} &  &  & X &  &  & X &  & X & X\tabularnewline
\cline{3-4} \cline{6-7} \cline{9-10} 
 &  & X & X &  & X &  &  & X & \tabularnewline
\cline{1-1} \cline{3-4} \cline{6-7} \cline{9-10} 
\multicolumn{1}{c}{} & \multicolumn{1}{c}{} & \multicolumn{1}{c}{} & \multicolumn{1}{c}{} & \multicolumn{1}{c}{} & \multicolumn{1}{c}{} & \multicolumn{1}{c}{} & \multicolumn{1}{c}{} & \multicolumn{1}{c}{} & \multicolumn{1}{c}{}\tabularnewline
\cline{1-1} \cline{3-4} \cline{6-7} \cline{9-10} 
\multirow{2}{*}{u-ch.} &  &  & X &  &  & X &  & X & X\tabularnewline
\cline{3-4} \cline{6-7} \cline{9-10} 
 &  & X & X &  & X & X &  & X & X\tabularnewline
\cline{1-1} \cline{3-4} \cline{6-7} \cline{9-10} 
\end{tabular}\caption{\label{tab:By-momentum-transfer}Forbidden combinations of energy
flow in all scattering-channel combinations of the overall T-channel
(and U-channel) by momentum transfer constraints.}
\end{table}

\noindent That is, for the overall T-channel (or for the overall
U-channel) eight possible configurations cannot yet be eliminated.

\subsection{Exclusion of t-channels in the overall T-channel\label{sub:Exclusion of t channels}}

The same argumentation that excluded the combination Sss in Chapter \ref{sub:Exclusion of Sss},
relying on a momentum transfer constraint, energy momentum conservation, and on the on-shellness of the 
loop modes, exclude the combinations Tts, Ttt, Ttu, Tst and Tut. Thus, we obtain an 
updated version of the exclusion Tab.\,\ref{tab:By-momentum-transfer} in terms of Tab.\,\ref{tab:By-momentum-transfer-and-enmomcons}. 

\begin{table}[H]
\centering{}%
\begin{tabular}{|c|c|c|c|c|c|c|c|c|c|}
\cline{1-1} \cline{3-4} \cline{6-7} \cline{9-10} 
\backslashbox{Vertex 2}{Vertex 1} &  & \multicolumn{2}{c|}{s-ch.} &  & \multicolumn{2}{c|}{t-ch.} &  & \multicolumn{2}{c|}{u-ch.}\tabularnewline
\cline{1-1} \cline{3-4} \cline{6-7} \cline{9-10} 
\multicolumn{1}{c}{} & \multicolumn{1}{c}{} & \multicolumn{1}{c}{} & \multicolumn{1}{c}{} & \multicolumn{1}{c}{} & \multicolumn{1}{c}{} & \multicolumn{1}{c}{} & \multicolumn{1}{c}{} & \multicolumn{1}{c}{} & \multicolumn{1}{c}{}\tabularnewline
\cline{1-1} \cline{3-4} \cline{6-7} \cline{9-10} 
\multirow{2}{*}{s- ch.} &  & X & X &  & X & X &  & X & X\tabularnewline
\cline{3-4} \cline{6-7} \cline{9-10} 
 &  & X & X &  & X & X &  &  & \tabularnewline
\cline{1-1} \cline{3-4} \cline{6-7} \cline{9-10} 
\multicolumn{1}{c}{} & \multicolumn{1}{c}{} & \multicolumn{1}{c}{} & \multicolumn{1}{c}{} & \multicolumn{1}{c}{} & \multicolumn{1}{c}{} & \multicolumn{1}{c}{} & \multicolumn{1}{c}{} & \multicolumn{1}{c}{} & \multicolumn{1}{c}{}\tabularnewline
\cline{1-1} \cline{3-4} \cline{6-7} \cline{9-10} 
\multirow{2}{*}{t-ch.} &  & X & X &  & X & X &  & X & X\tabularnewline
\cline{3-4} \cline{6-7} \cline{9-10} 
 &  & X & X &  & X & X &  & X & X\tabularnewline
\cline{1-1} \cline{3-4} \cline{6-7} \cline{9-10} 
\multicolumn{1}{c}{} & \multicolumn{1}{c}{} & \multicolumn{1}{c}{} & \multicolumn{1}{c}{} & \multicolumn{1}{c}{} & \multicolumn{1}{c}{} & \multicolumn{1}{c}{} & \multicolumn{1}{c}{} & \multicolumn{1}{c}{} & \multicolumn{1}{c}{}\tabularnewline
\cline{1-1} \cline{3-4} \cline{6-7} \cline{9-10} 
\multirow{2}{*}{u-ch.} &  &  &  &  & X & X &  & X & X\tabularnewline
\cline{3-4} \cline{6-7} \cline{9-10} 
 &  & X & X &  & X & X &  & X & X\tabularnewline
\cline{1-1} \cline{3-4} \cline{6-7} \cline{9-10} 
\end{tabular}\caption{\label{tab:By-momentum-transfer-and-enmomcons}Forbidden combinations of energy
flow in all scattering-channel combinations of the overall T-channel
(and U-channel) by momentum transfer constraints and energy momentum conservation.}
\end{table}
Again, only four configurations are left unexcluded.

\subsection{Monte-Carlo analysis of remaining cases\label{sub:Monte-Carlo-analysis-of}}

For 4 out of the 36 configurations in each of the overall S-, T-, and U-channels no analytical
exclusion could be performed. These remaining cases thus are treated numerically. To obtain an estimate on 
numerical precision in sampling the according algebraic varieties in these non-excluded combinations we 
also sample the analytically excluded Sss configuration. To proceed, 
a suitable set of non-redundant variables must be defined. As we will see, 
it is possible to parametrize the overall scattering process by the
variables (referring to Fig.\,\ref{fig:All-3-Overall-Ch.}) $a_{0}$,
$b_{0}$, the energies of the incoming photons, the angle $\alpha\equiv\measuredangle\mathbf{a}\mathbf{b}\in\left[0,2\pi\right]$
between their three-momenta, and the angles $\zeta\in\left[-\frac{\pi}{2},\frac{\pi}{2}\right]$
and $\eta\in\left[0,2\pi\right]$, that are necessary to describe
the spatial orientation of one of the outgoing particles. Moreover
two angles, $\theta\in\left[-\frac{\pi}{2},\frac{\pi}{2}\right]$ and
$\varphi\in\left[0,2\pi\right]$, are sufficient to describe the kinematic state
of the internal particles. 
\noindent 
\begin{figure}[H]
\begin{centering}
\includegraphics[width=0.48\textwidth]{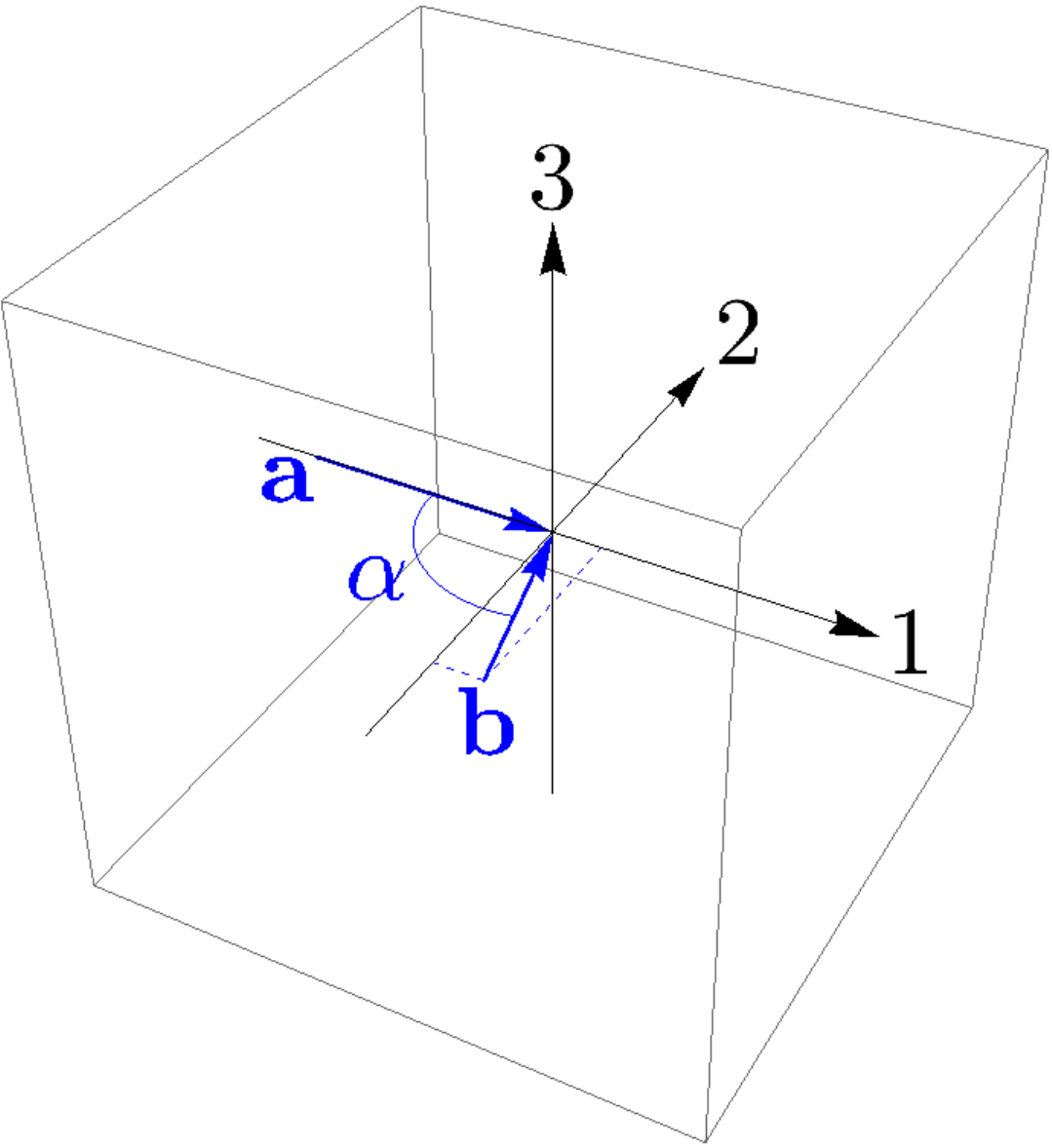}$\,$\includegraphics[width=0.48\textwidth]{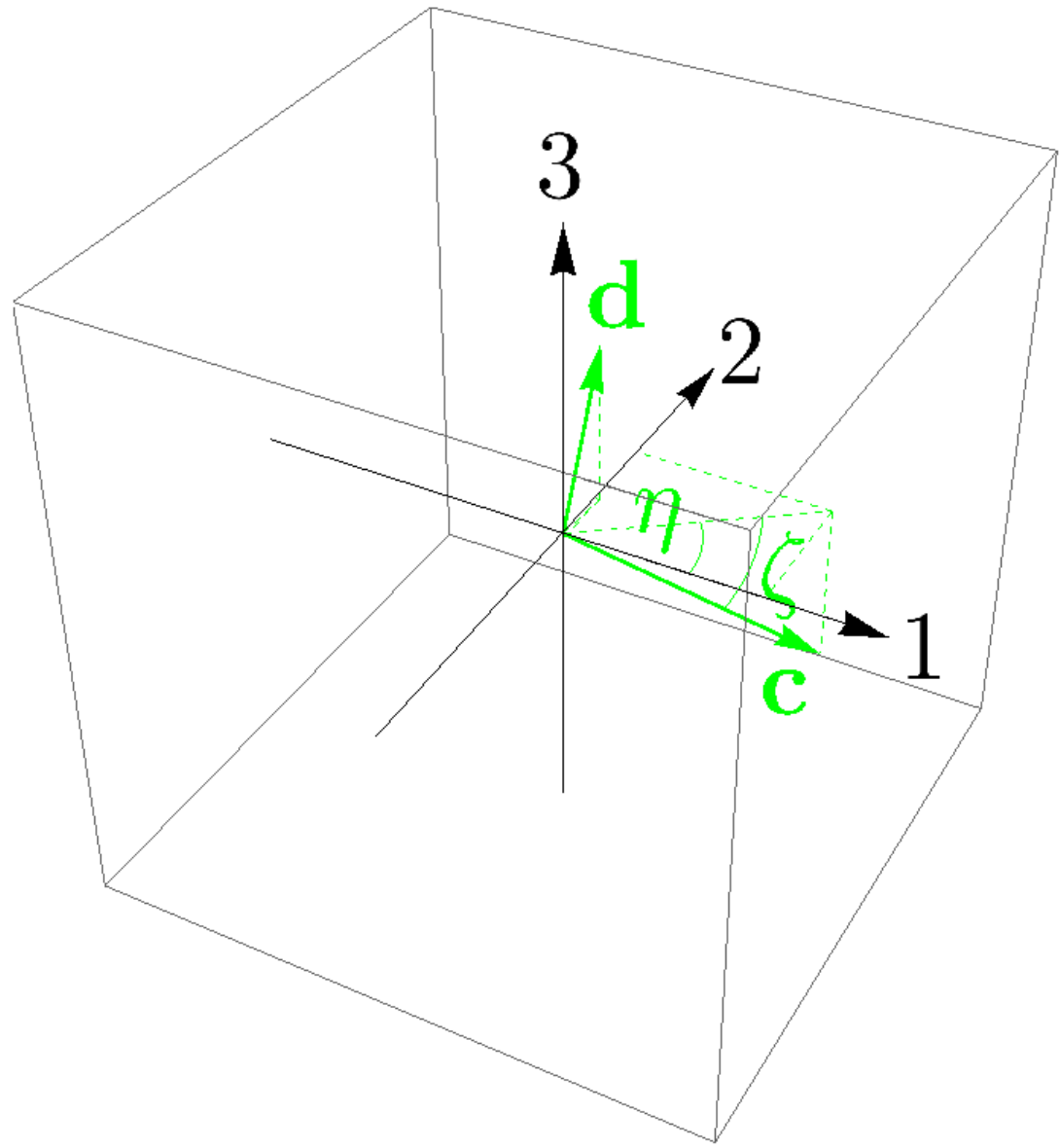}
\par\end{centering}

\begin{centering}
\includegraphics[width=0.48\textwidth]{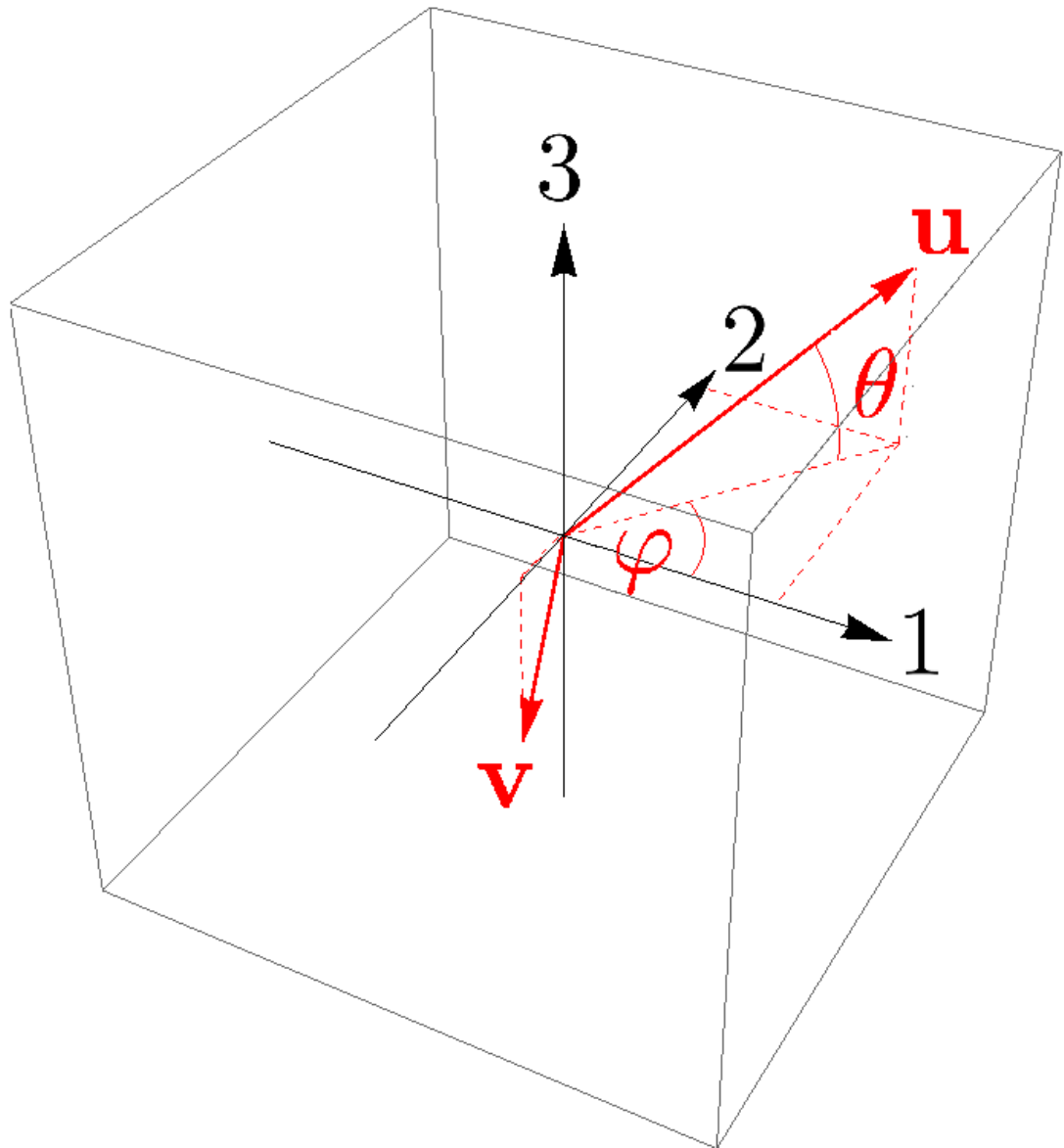}
\par\end{centering}

\caption{A visualization of the involved three-momenta (redundant variables).
The incoming momenta are blue, the internal ones are red, and the outgoing
momenta are green.}
\end{figure}

\noindent In the following we relate the involved four-momenta with
respect to these parameters. First, we exploit four-momentum conservation and 
that the external photons are on shell:

\noindent 
\begin{align}
 & a=\left(\begin{array}{c}
a_{0}\\
a_{0}\\
0\\
0
\end{array}\right)\:,\: b=\left(\begin{array}{c}
b_{0}\\
b_{0}\cos\left(\alpha\right)\\
b_{0}\sin\left(\alpha\right)\\
0
\end{array}\right)\:,\: c=\left(\begin{array}{c}
c_{0}\\
c_{0}\cos\left(\zeta\right)\cos\left(\eta\right)\\
c_{0}\cos\left(\zeta\right)\sin\left(\eta\right)\\
c_{0}\sin\left(\zeta\right)
\end{array}\right)\:,\: d=a+b-c\label{eq:abcd}
\end{align}
Eq.\,(\ref{eq:abcd}) implies 
\noindent 
\begin{align*}
 & 0=d^{2}=\left(a+b-c\right)^{2}=2ab-2ac-2bc\\
 & \phantom{0}=2\left(a_{0}b_{0}-\left(a_{0}+b_{0}\right)c_{0}-a_{0}b_{0}\cos\left(\alpha\right)+c_{0}\left(b_{0}\cos\left(\alpha-\eta\right)+a_{0}\cos\left(\eta\right)\right)\sin\left(\zeta\right)\right)\\
 & \Rightarrow c_{0}=\frac{a_{0}b_{0}(1-\cos(\alpha))}{a_{0}+b_{0}-b_{0}\cos\left(\alpha-\eta\right)\sin\left(\zeta\right)-
a_{0}\cos\left(\eta\right)\sin\left(\zeta\right)}\,,
\end{align*}
independently of the overall scattering channel S, T, or U. Momenta of the massive modes, however, are dependent
on S, T, or U. Both internal particles are on shell. For the overall S-channel and for given incoming momenta $\tilde{a},\tilde{b}$ 
the absolute value of loop three-momentum $\left|\mathbf{u}\right|$ is related to its orientation $\mathbf{e_{u}}$, 
which depends on $\theta$ and $\varphi$, as given in Eq. \eqref{eq:iniLsg1/2}. 
We have
\noindent 
\begin{align}
\left|\mathbf{\tilde{u}}\right|_{1/2}^{S}= & \frac{\left(\tilde{a}\tilde{b}\right)\left(\mathbf{\tilde{a}}+\mathbf{\tilde{b}}\right)\mathbf{e_{u}}}{\left(\tilde{a}_{0}+\tilde{b}_{0}\right)^{2}-2\left(\left(\mathbf{\tilde{a}}+\mathbf{\tilde{b}}\right)\mathbf{e_{u}}\right)^{2}}\nonumber \\
 & \pm\frac{\left(\tilde{a}_{0}+\tilde{b}_{0}\right)\sqrt{-\tilde{m}^{2}\left(\tilde{a}_{0}+
\tilde{b}_{0}\right)^{2}+\left(\tilde{a}\tilde{b}\right)^{2}+\tilde{m}^{2}\left(\left(\mathbf{\tilde{a}}+
\mathbf{\tilde{b}}\right)\mathbf{e_{u}}\right)^{2}}}{\left(\tilde{a}_{0}+\tilde{b}_{0}\right)^{2}-
\left(\left(\mathbf{\tilde{a}}+\mathbf{\tilde{b}}\right)\mathbf{e_{u}}\right)^{2}}\,.\label{eq:iniLsg1/2-1}
\end{align}
\noindent The expressions for the overall T- and U-channels can be
obtained by interchanging momenta: The T-channel relates to the
S-channel by exchanging $a\rightarrow c$ and $b\rightarrow-a$; the
U-channel relates to the S-channel by $a\rightarrow d$ and $b\rightarrow-a$. Clearly, the
energy $u_{0}$ is determined by on-shellness, and the other internal four-momentum $\tilde{v}$
is completely determined by four-momentum conservation. Whether or not a given set
of parameter values satisfies all constraints is tested by inserting it into Eqs. \eqref{eq:abcd}
and \eqref{eq:iniLsg1/2-1} before, in turn, the 4-vertex constraints are probed at a given value of the 
dimensionless temperature $\lambda$. Because $\lambda$, $a_{0}$, $b_{0}$ are not bounded from above 
their to-be-tested values need to be limited in the numerical procedure. We require $\lambda_c=13.867\le\lambda\le 100$ \cite{Hofmann2005} 
and $\hat{a}_{0}=\frac{a_{0}}{T}$, $\hat{b}_{0}=\frac{b_{0}}{T}\leq 100$. Here $\lambda_c$ is the critical temperature for the deconfining-preconfining 
phase transition where $m\to\infty$ and massive modes thus decouple. 

Parameter values are sampled randomly in a conditioned way, 
and parameter sets satisfying the constraints are counted (for detailed information, see Appendix \ref{sec:Appendix}).

\subsubsection{Typical hit densities\label{hitdens}}

\begin{figure}[H]
\centering{}\includegraphics[width=0.8\textwidth]{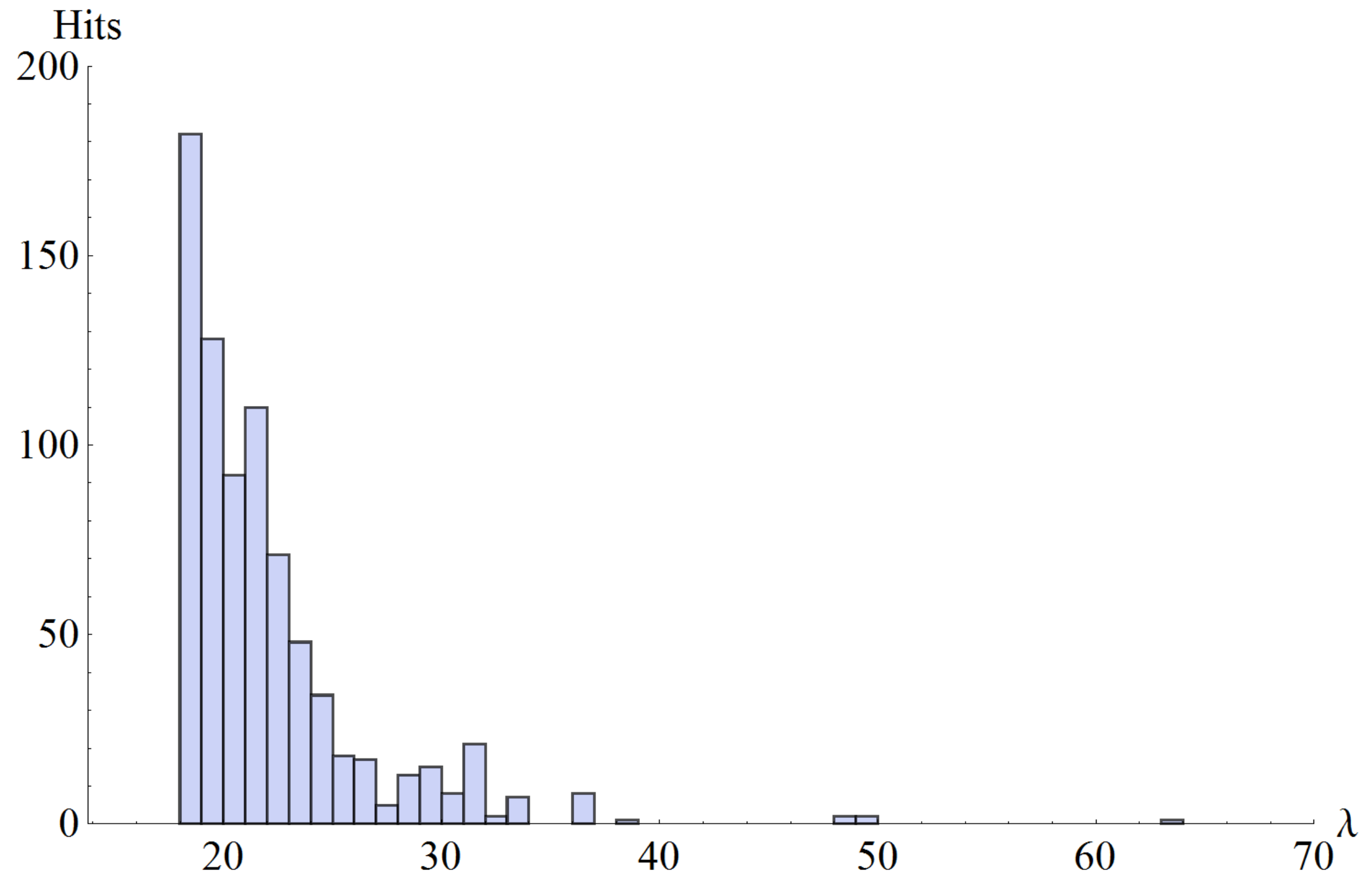}\caption{\label{fig:LambdaHisto}
This histogram shows the distribution of the sum of hits over all tested channel combinations as a function of dimensionless
temperature $\lambda$ in the range from $\lambda_{c}=13.867$ to $100$.}
\end{figure}

In the overall S channel, 87 out of $6.144\times10^{10}$ tested
parameter sets satisfied the constraints. This number is suppressed by a factor $\sim$ thirteen  
compared to the overall T-channel where $1110$ out of $6.144\times10^{10}$ sets satisfied the constraints. Because
of the afore mentioned symmetry of the constraints under the exchange of outgoing photons T- and U-channel should yield identical results. 
The number quoted for the T-chanell represents the average of
all four remaining combinations in Tsu and Tus. We conclude from the S- versus T- plus U- 
channel comparison that the former is supported by less than 5\% of the integration volume of the latter two channels. 
Practically, this excludes the overall S-channel. We also investigated 
the distribution of valid parameter values. For example, Fig. \ref{fig:LambdaHisto} depicts the abundance of the sum of 
hits over all non-excluded channel combinations as a function of $\lambda$.
\noindent 
\noindent Two things are worth pointing out. First, we see that processes
with $\lambda>40$ are very rare. The highest temperature associated with a valid 
configuration is $\lambda=63.32$, and this is an extreme outlier. Second, the
abundance is rapidly decaying for $\lambda \le 18.8$. In fact, a contribution at a temperature smaller than 
$\lambda=18.15$ never was detected. 

\begin{figure}[H]
\begin{centering}
\includegraphics[width=1\textwidth]{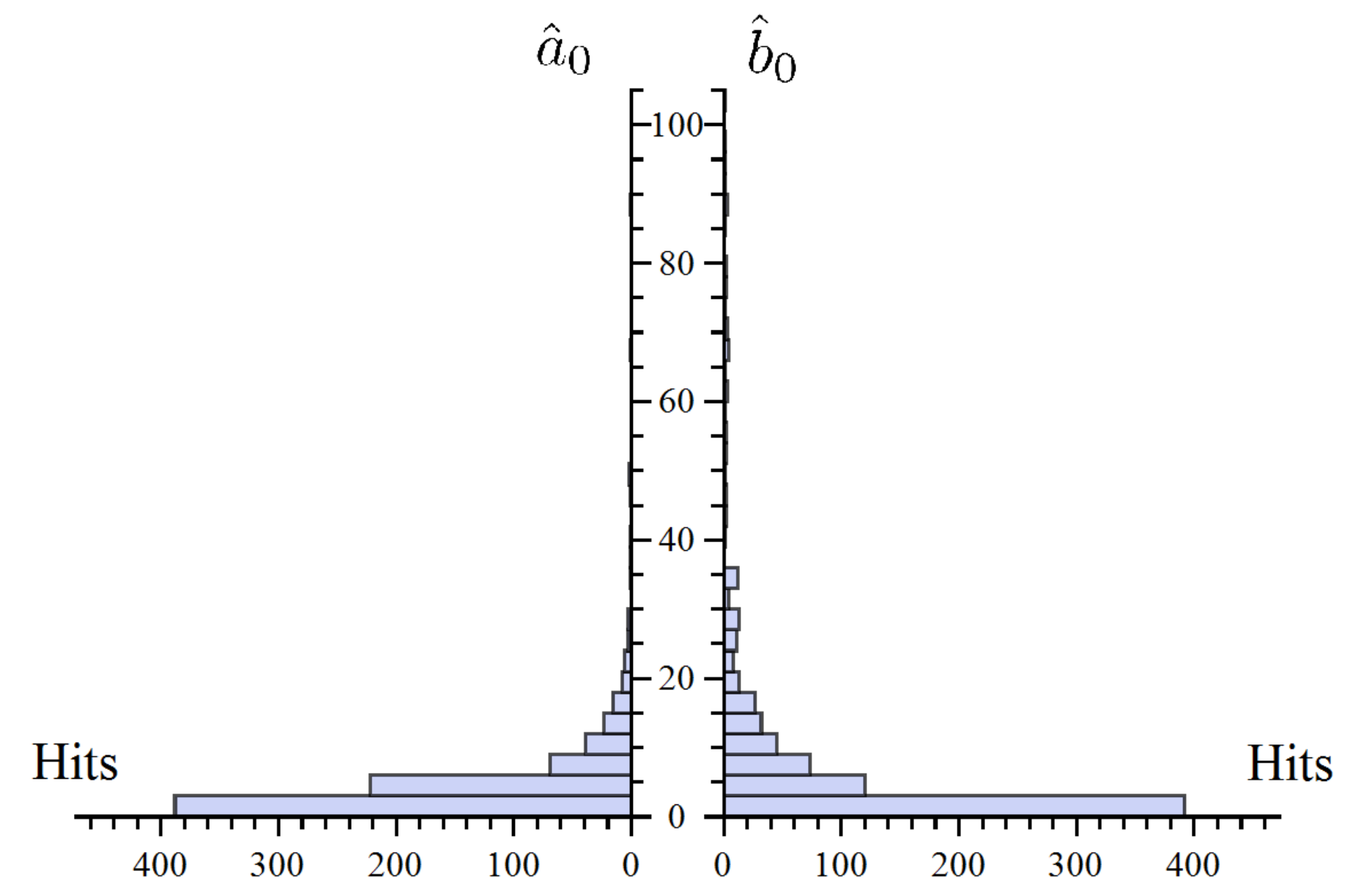}
\par\end{centering}
\caption{\label{fig:A0B0Histo}Comparison of the distributions of the two incoming 
energies, normalized with respect to temperature $T$. Hits in each bin 
represent the sum over all tested channel combinations.}
\end{figure}

Another interesting distribution is the abundance of 
energies relative to temperature $T$, $\hat{a}_{0}$ and $\hat{b}_{0}$, as shown in Fig. \ref{fig:A0B0Histo}. 
As anticipated, there is no obvious difference between the distributions of $\hat{a}_{0}$
and $\hat{b}_{0}$. The constraints seem to imply an upper bound of about $40$. Configurations with values 
of $\hat{a}_{0}$ and $\hat{b}_{0}$ above this bound are anyway 
strongly suppressed by the Bose-Einstein distributions associated with external, thermalized photons.  

\subsubsection{Algebraic varieties\label{algvar}}

\begin{figure}[H]
\begin{centering}
\includegraphics[width=0.45\textwidth]{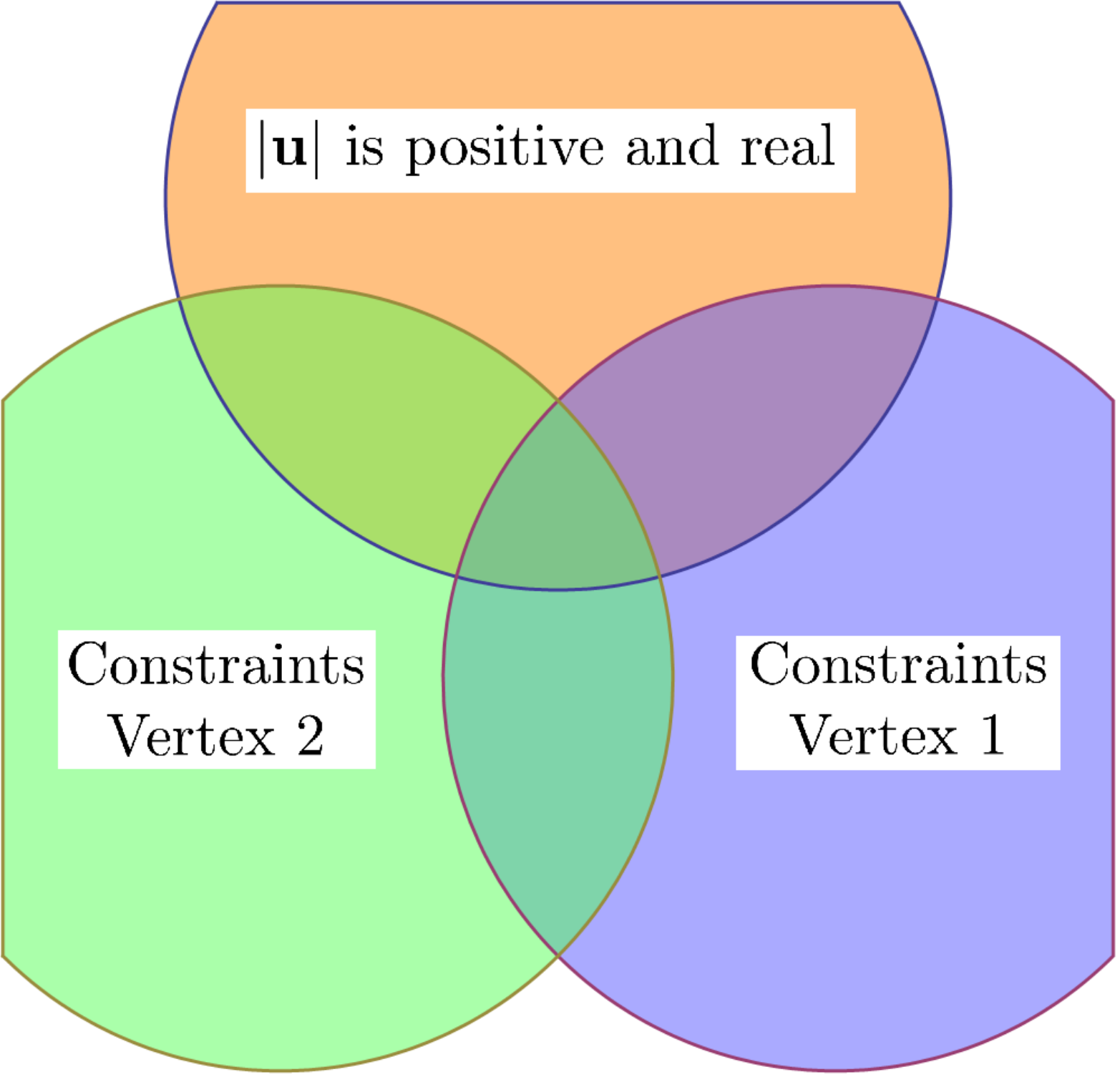}
\par\end{centering}
\caption{\label{fig:Color Codification}This region plot depicts 
color coded representations of the effects of single and combined constraints. The varieties marked in {\sl orange} are associated with real 
and positive solutions $\left|\mathbf{u}\right|$ to Eq. \eqref{eq:iniLsg1/2-1}. 
Varieties, where the momentum transfer constraints in the first vertex
are satisfied, are marked in {\sl green}, and those, where the momentum
transfer constraints are fulfilled at the second vertex, are indicated in {\sl blue}. 
The finally valid variety is represented by the intersection of these three varieties.}
\end{figure}

\begin{figure}[H]
\centering{}%
\begin{tabular}{cc}
\includegraphics[width=0.37\textwidth,height=0.37\textwidth]{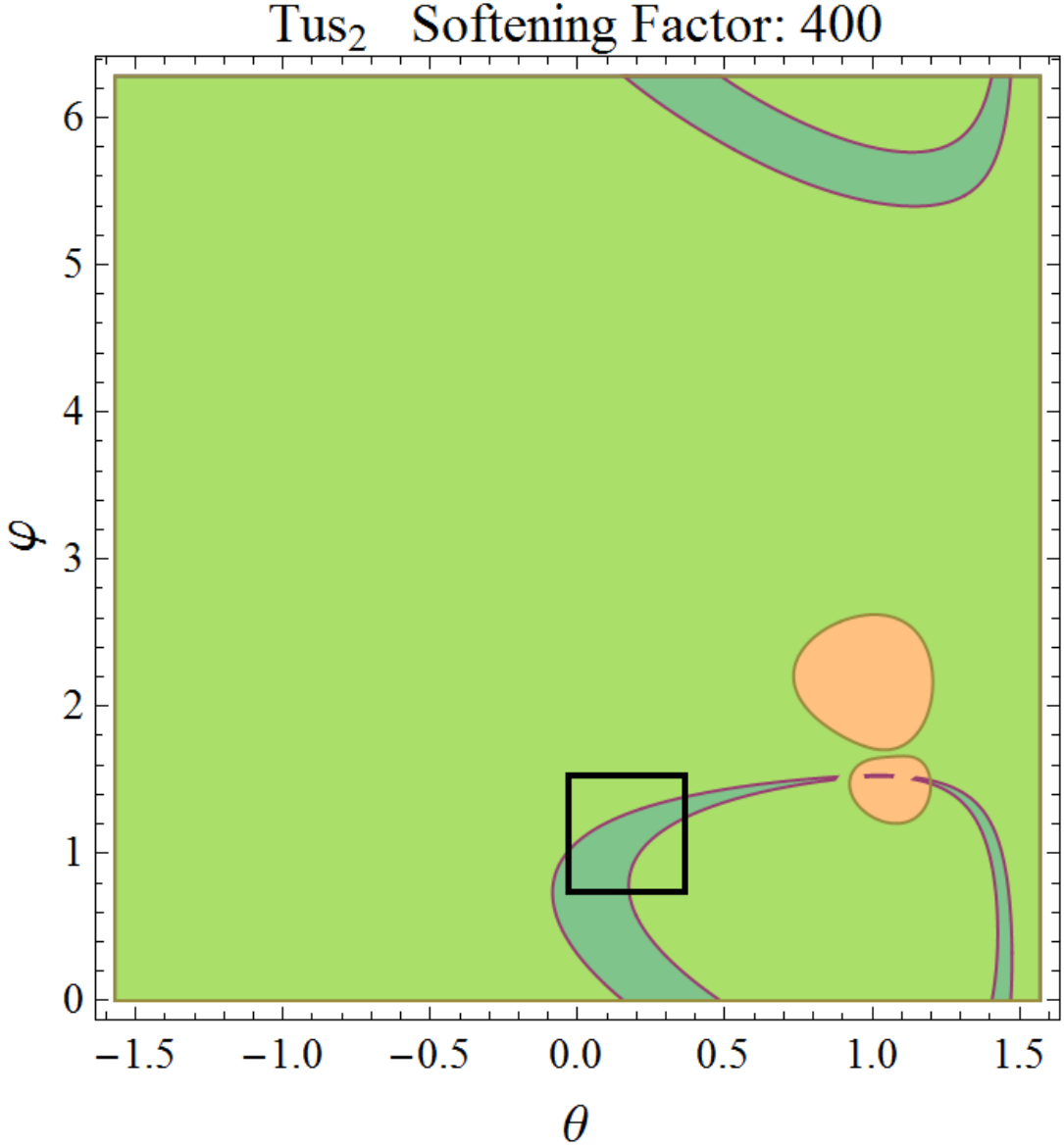} & \hspace{-9 pt}\includegraphics[width=0.37\textwidth,height=0.37\textwidth]{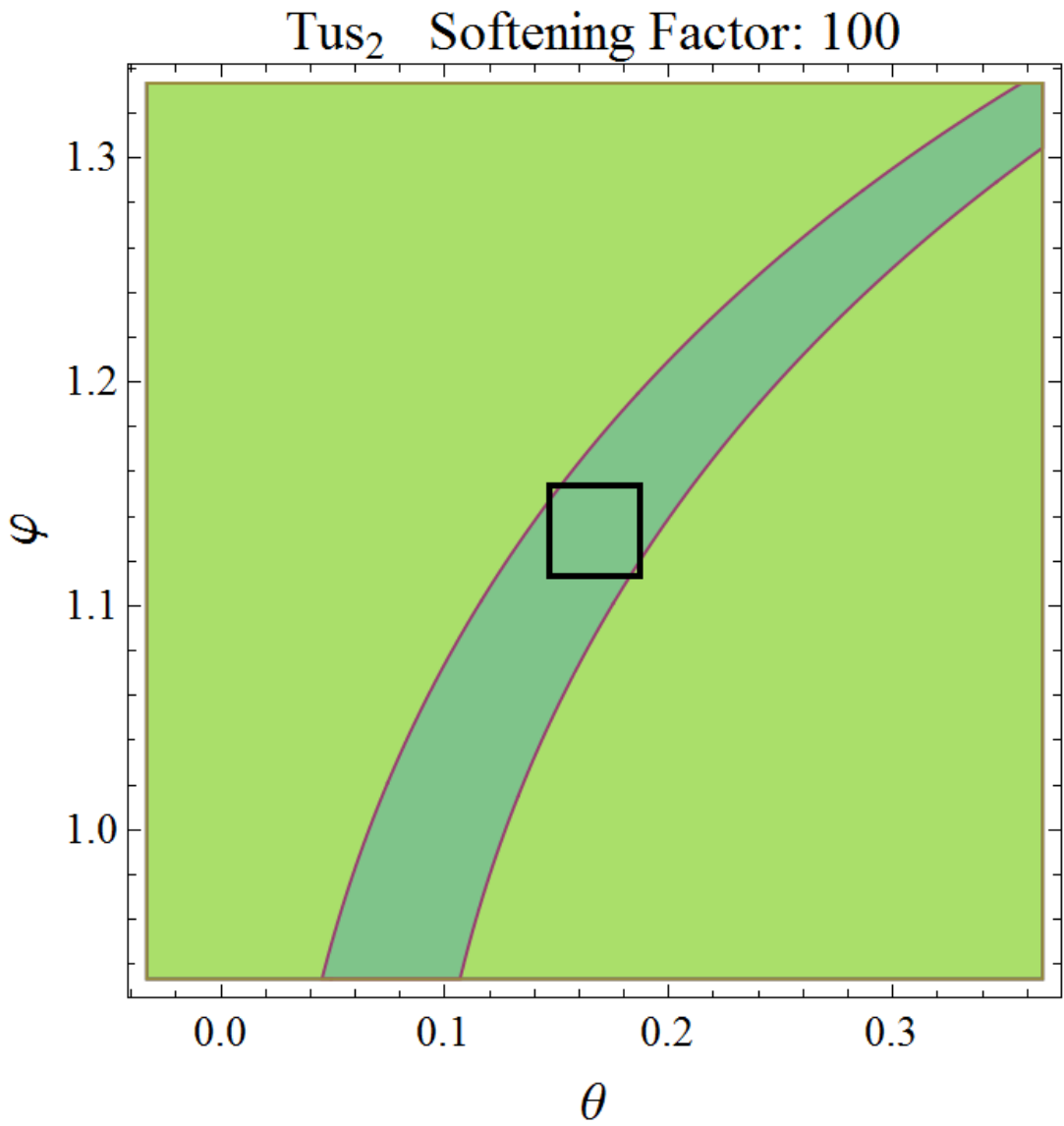}\tabularnewline
\includegraphics[width=0.37\textwidth,height=0.37\textwidth]{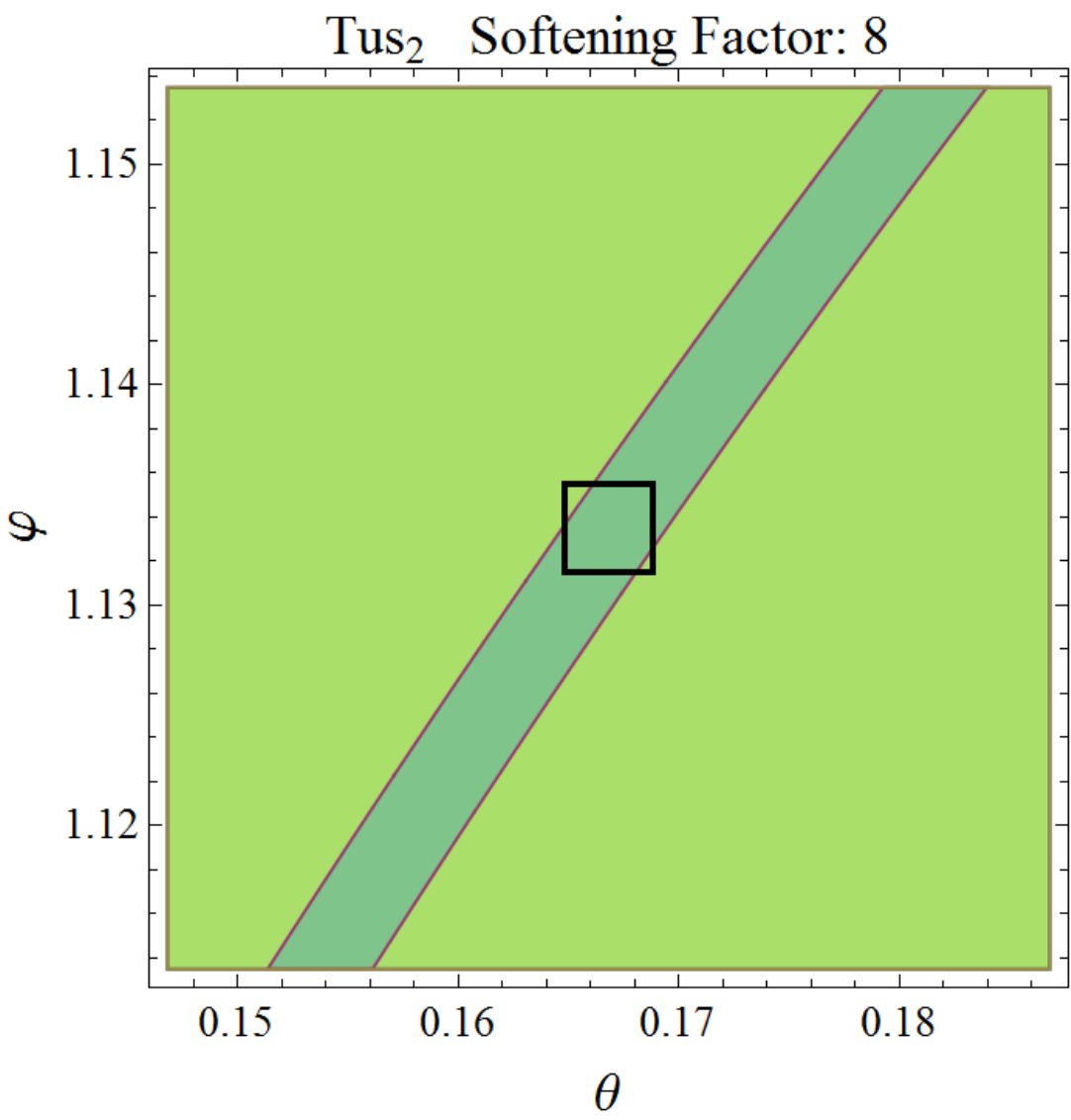} & \hspace{-9 pt}\includegraphics[width=0.37\textwidth,height=0.37\textwidth]{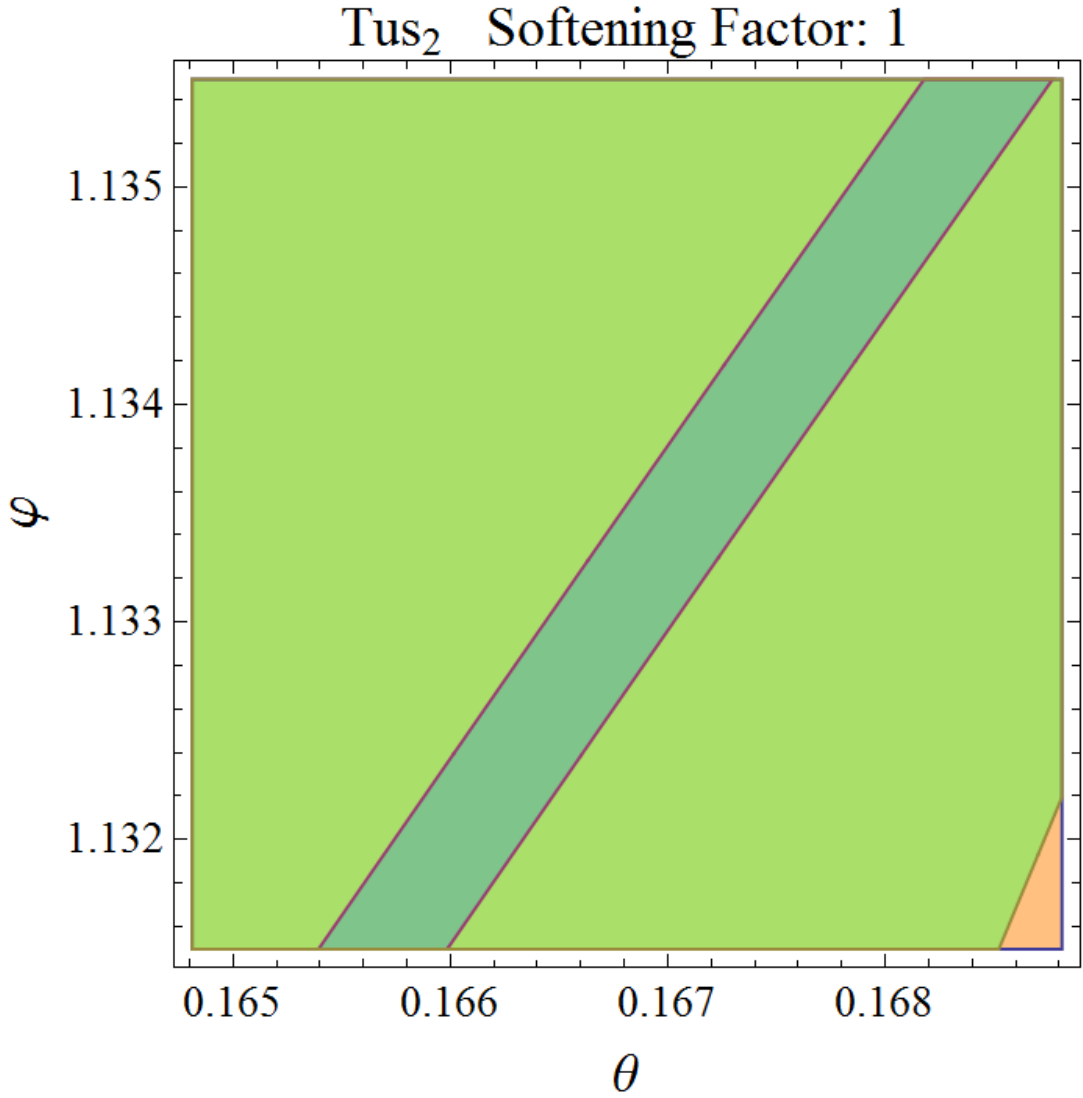}\tabularnewline
\end{tabular}\caption{\label{fig:Tut1}A visualization of the constraints in the $\theta$-$\varphi$
plane in the overall T-channel with scattering channels u and
s for $\mbox{\ensuremath{\left|\mathbf{u}\right|}}_{2}$ and the following parameter values of the pivotal 
configuration: $\lambda=25.727$, $\tilde{a}_{0}=36.5239$, $\tilde{b}_{0}=11.0003$,
$\alpha=1.53077$, $\eta=5.82265$, $\zeta=2.09548$.}
\end{figure}

To explore the numerical data further, we zoom into the $\theta$-$\varphi$
plane about valid parameter-value combinations 
($\lambda\mbox{, }\alpha\mbox{, }\beta\mbox{, }\gamma\mbox{, }\tilde{a}_{0}\mbox{ and }\tilde{b}_{0}$). It was numerically not 
possible to resolve the associated varieties of valid configuration about a particular one at once. Therefore we imposed a relaxation
of the constraints to broaden the region of valid configurations. The relaxation is implemented
by virtue of a softening factor $\Upsilon$ implemented as $1\rightarrow\Upsilon$ on the
right hand sides in Eqs. \eqref{eq:CombisAngang} to \eqref{eq:CombisEnde-1}. 
(The momentum transfer must only be smaller than $\Upsilon\left|\phi\right|^{2}$ in the relaxed as opposed to the physical situation.) 
For the combination shown in Fig.\,\ref{fig:Tut1}, 
representative of all other non-excluded combinations, there are series of four region plots. The first (top left) depicts 
a region of interest of size $\Delta\theta=0.4$ and $\Delta\varphi=0.8$ which is centered around the pivotal hit detected during
the Monte Carlo test. This region is blown up into the second plot
(top right) where, in turn, a centered region of interest of size $\Delta\theta=0.04$
and $\Delta\varphi=0.08$ is shown. Again, the latter is blown up into the third region plot (bottom left). 
The region of interest marked here and shown in full size in plot four (bottom right) has an extent of $\Delta\theta=0.004$
and $\Delta\varphi=0.008$. This last plot (lower right) represents the 
physical situation with $\Upsilon=1$. Values of the softening factor $\Upsilon$ are chosen to point out the nature of the 
actual, physical variety. We need to distinguish the two realizations for each channel combination
(corresponding to the $\pm$ in the expression for $\left|\mathbf{u}\right|$
in Eq. \eqref{eq:iniLsg1/2-1}). The solution corresponding to $+$
is labeled by the index $1$, and the solution corresponding to the
minus sign has the index $2$. The contour plots in the $\theta$-$\varphi$ plane signal the different
constraints by a color code as explained in the caption
of Fig. \ref{fig:Color Codification}.\\


\noindent 

\noindent 

\section{Summary and Conclusions}

In the present work we have discussed the nature of non-thermalities in SU(2) Yang-Mills theory, defined by a functional 
integral involving periodic gauge-field configurations. While the plane-wave sector behaves entirely 
thermal, it is the sector of isolatedly acting topological field configurations, which bears the seed of 
non-thermal behavior. Furthermore, we have investigated systematically how the unitary-gauge constraints of the effective theory for 
deconfining SU(2) Yang-Mills thermodynamics limit the contributions of loop momenta to 
the amplitude for one-loop photon-photon scattering. Only one type of Feynman diagrams with two 4-vertices is admissible 
to mediate this process, and a large part of channel and energy-sign combinations for the scattering through 
these vertices is analytically excluded relying on a subset of all energy-flow and momentum-transfer constraints. Out of a total of 
108 scattering-channel and energy-sign configurations for the internal modes 12 configurations 
cannot be excluded analytically. These remaining cases do not give rise to pair creation or annihilation (practically, there is an exclusion of the overall S-channel) 
and were analyzed by Monte-Carlo sampling subject to all constraints. The associated hit densities decay very rapidly 
with temperature. We have also investigated the admissible algebraic variety in the vicinity of a selected, pivotal Monte-Carlo 
hit to demonstrate how filamentous it is.

One may, at first sight, object that an analysis of allowed regions for the loop integration is not sufficient to draw a 
conclusion about the actual smallness of the integral for the amplitude since singular integrands may arise on
the filamentous integration regions. This is not the case, however, because we consider a reduced integration manifold,  
obtained after singular distributions in the original integrand are integrated out. 
(On-shell conditions, associated with $\delta$-distributions in the full 4$l$-dimensional space of loop integration at loop order $l$, 
are implemented in the analysis of vertex conditions from the start. Thus the integration manifold considered here  
is lower dimensional than 4$l$. Integrands are either regular, or if singular, thanks to their integrability  
can be made regular by non-singular changes of variables.)          

All in all, our results suggest that photon-photon scattering within the deconfining SU(2) Yang-Mills plasma is 
feeble: Practically, it does not occur in the overall S-channel, 
excluding the creation of massive modes out of photons and vice versa by the optical theorem. This is 
in agreement with experiment\footnote{In \cite{Hofmann2011} it is explained at 
length why the gauge-group factor U(1)$_Y$ of the present Standard Model of Particle Physics should be interpreted 
as the Cartan group of dynamically broken SU(2) Yang-Mills theory of scale $\Lambda\sim 10^{-4}\,$eV.}. 

Low-temperatures modification in thermal photon propagation from conventional U(1) behaviour, 
which could explain the large-angle anomalies of observed temperature fluctuations in the Cosmic Microwave 
Background \cite{Hofmann2013}, is largely dominated by the photon polarisation tensor. However, 
it is well possible that the feeble one-loop photon-photon correlation introduced by photon-photon 
scattering at low temperatures affects the Cosmic Microwave Background's polarisation at low redshift 
(the domain of loop integration represents less than $10^{-7}$ times the volume of the  
unconstrained integration region). More work is required to actually match 
this to observational results \cite{Bicep2}.  

Finally, a higher dimensional generalization (more than two 
4-vertices per diagram) of the technology of energy-flow exclusion developed in the present 
work may be key to proving the termination of the expansion of thermodynamical quantities into 
irreducible bubble diagrams at a finite loop order, conjectured in \cite{Hofmann2006} and based on a counting of constraints 
versus independent radial loop variables in dependence of loop order.

\section{Appendix\label{sec:Appendix}}

\noindent Here we sketch the sampling strategy referred to in Sec.\,\ref{sub:Monte-Carlo-analysis-of}.
Because we would like to do justice to the different scales of the
sampling ranges and since runtime resources were limited we nested the sampling of random variables 
and required that noncompact variables are tested much more 
often than compact ones. The algorithm is sketched below in terms of a pseudocode:\\
$\,$\vspace{0.5cm}\\
repeat 60 times:

\noindent $\qquad$$\lambda=$ random real number $\in\left[13.867,100\right]$;

\noindent $\qquad$repeat 4 times:

\noindent $\qquad$$\qquad$$\alpha=$ random real $\in\left[0,2\pi\right]$;

\noindent $\qquad$$\qquad$repeat 4 times:

\noindent $\qquad$$\qquad$$\qquad$$\eta=$ random real $\in\left[0,2\pi\right]$;

\noindent $\qquad$$\qquad$$\qquad$repeat 2 times:

\noindent $\qquad$$\qquad$$\qquad$$\qquad$$\zeta=$ random real
$\in\left[-\frac{\pi}{2},\frac{\pi}{2}\right]$;

\noindent $\qquad$$\qquad$$\qquad$$\qquad$repeat 2 times:

\noindent $\qquad$$\qquad$$\qquad$$\qquad$$\qquad$$\theta=$
random real $\in\left[-\frac{\pi}{2},\frac{\pi}{2}\right]$;

\noindent $\qquad$$\qquad$$\qquad$$\qquad$$\qquad$repeat 4 times:

\noindent $\qquad$$\qquad$$\qquad$$\qquad$$\qquad$$\qquad$$\varphi=$
random real $\in\left[0,2\pi\right]$;

\noindent $\qquad$$\qquad$$\qquad$$\qquad$$\qquad$$\qquad$repeat
$10^{4}$ times:

\noindent $\qquad$$\qquad$$\qquad$$\qquad$$\qquad$$\qquad$$\qquad$$\hat{a}_{0}=$
random real $\in\left[0,100\right]$;

\noindent $\qquad$$\qquad$$\qquad$$\qquad$$\qquad$$\qquad$$\qquad$$\hat{b}_{0}=$
random real $\in\left[0,100\right]$;

\noindent $\qquad$$\qquad$$\qquad$$\qquad$$\qquad$$\qquad$$\qquad$test
if constraints for given channel-combination

\noindent $\qquad$$\qquad$$\qquad$$\qquad$$\qquad$$\qquad$$\qquad$
are satisfied for the given parameters 

\noindent $\qquad$$\qquad$$\qquad$$\qquad$$\qquad$$\qquad$$\qquad$$\lambda$,
$\hat{a}_{0}$, $\hat{b}_{0}$, $\alpha$, $\eta$, $\zeta$, $\theta$,
$\varphi$;

\noindent $\qquad$$\qquad$$\qquad$$\qquad$$\qquad$$\qquad$$\qquad$if
yes:

\noindent $\qquad$$\qquad$$\qquad$$\qquad$$\qquad$$\qquad$$\qquad$$\qquad$increase
hitcount by one;

\noindent $\qquad$$\qquad$$\qquad$$\qquad$$\qquad$$\qquad$$\qquad$$\qquad$save
parameters;

\end{document}